\documentclass[aps,preprint,endfloats*]{revtex4}
\usepackage{graphicx,epsfig,graphics,amssymb,amsmath,subeqnarray}

\def\np{\nabla_{\perp}}
\def\e{\epsilon}
\def\p{\partial}
\def\be{\begin{equation}}
\def\ee{\end{equation}}
\def\d{{\rm d}}
\def\besub{\begin{subeqnarray}}
\def\eesub{\end{subeqnarray}}

\begin{document}

\title{Three-dimensional flows in slowly-varying planar geometries}
\author{Eric Lauga$^1$, Abraham D. Stroock$^2$ and Howard A. Stone$^1$}
\affiliation{$^1$Division of Engineering and Applied Sciences,
Harvard University, Cambridge, MA 02138,\\
$^2$School of Chemical and Biomolecular Engineering, Cornell
University, Ithaca, NY 14853.}
\date{\today}

\begin{abstract}
We consider laminar flow in channels constrained geometrically to
remain between two parallel planes; this geometry is typical of
microchannels obtained with a single step by current
microfabrication techniques. For pressure-driven Stokes flow in
this geometry and assuming that the channel dimensions change
slowly in the streamwise direction, we show that the velocity
component perpendicular to the constraint plane cannot be zero
unless the channel has both constant curvature and constant
cross-sectional width. This result implies that it is, in
principle, possible to design ``planar mixers", {\it i.e.} passive
mixers for channels that are constrained to lie in a flat layer
using only streamwise variations of their in-plane dimensions.
Numerical results are presented for the case of a channel with
sinusoidally varying width.


\end{abstract}
\maketitle

\section{Introduction}

The rapid development of microfluidic systems and their
applications in domains such as aeronautics, chemistry, material
synthesis, medical diagnostics and drug delivery has recently
motivated investigations of new research questions in fluid
dynamics at small scales (\cite{HoRev,Whitesides,Stone}). A
particularly active research area is the design of mixers for
microdevices. This task has proven itself to be a real engineering
challenge due to a variety of physical and practical constraints.
Physically, the typical cross-sectional dimensions of
microchannels in current microfluidic systems are on the order of
$10-100~\mu$m (\cite{Manz,Whitesides}). On this scale, practical
flows  of common liquids ($U\approx 0.1-10$ cm/s) have low
Reynolds numbers, $Re < 10$, and turbulent mixing of the fluid
does not generally occur.  At the same time, the Peclet number for
mass transfer in these flows, defined as the ratio of a typical
diffusive time to a typical advection time is high, $Pe
> 100$, and purely diffusive mixing across the flow is slow.
For applications that require mixing, it is therefore necessary to
design channels that will lead to efficient convective mixing.

In a channel geometry, the strongest gradients of concentration
are typically oriented in a direction normal to the principal axis
of the channel, because the gradients exist between co-flowing
streams of distinct chemical makeup.  An effective mixing flow in
a channel must therefore stir the fluid over the cross-section
with transverse flows as the fluid progresses downstream in the
axial flow; such a flow must possess three non-zero components. An
effective stirring flow rapidly folds the fluid into itself so as
to decrease the distance that diffusion must act to homogenize
concentrations. The potential of a given channel-flow to mix
efficiently can be judged by several characteristics:  (1) the
ratio of the transverse to the axial velocities.  If this ratio is
too small, then the axial length of channel required for mixing is
likely to be impractically long, regardless of the detailed
character of the flow;   (2) the distribution of transverse flows
within the cross-section of the channel.  If the flows are
confined to small areas within the cross-section, then they will
be ineffective at exchanging fluid between these areas and with
areas in which no flow exists.  (3) the evolution of the
transverse flows as a function of axial position along the
channel.  An efficient mixer should produce Lagrangian chaos; in
particular, no streamwise symmetries should be present \cite{sym}.
This feature is achieved in a channel-geometry when the position
and orientation of transverse flows vary axially such that all
fluid elements travelling in the channel experience an alternating
sequence of rotational and extensional flows (\cite{Ottino}).

In designing a laminar mixer for microfluidic applications, it is
also important to take into account the constraints imposed by
technology and conversation laws. Designs of mixers should be
scalable to smaller systems, resistant to fouling by particulate
matter, and efficient with respect to power consumption. Moreover,
the typical lithographic processes currently used in
microfabrication lead to planar geometries.

\begin{figure}
\centering
\includegraphics[width=.8\textwidth]{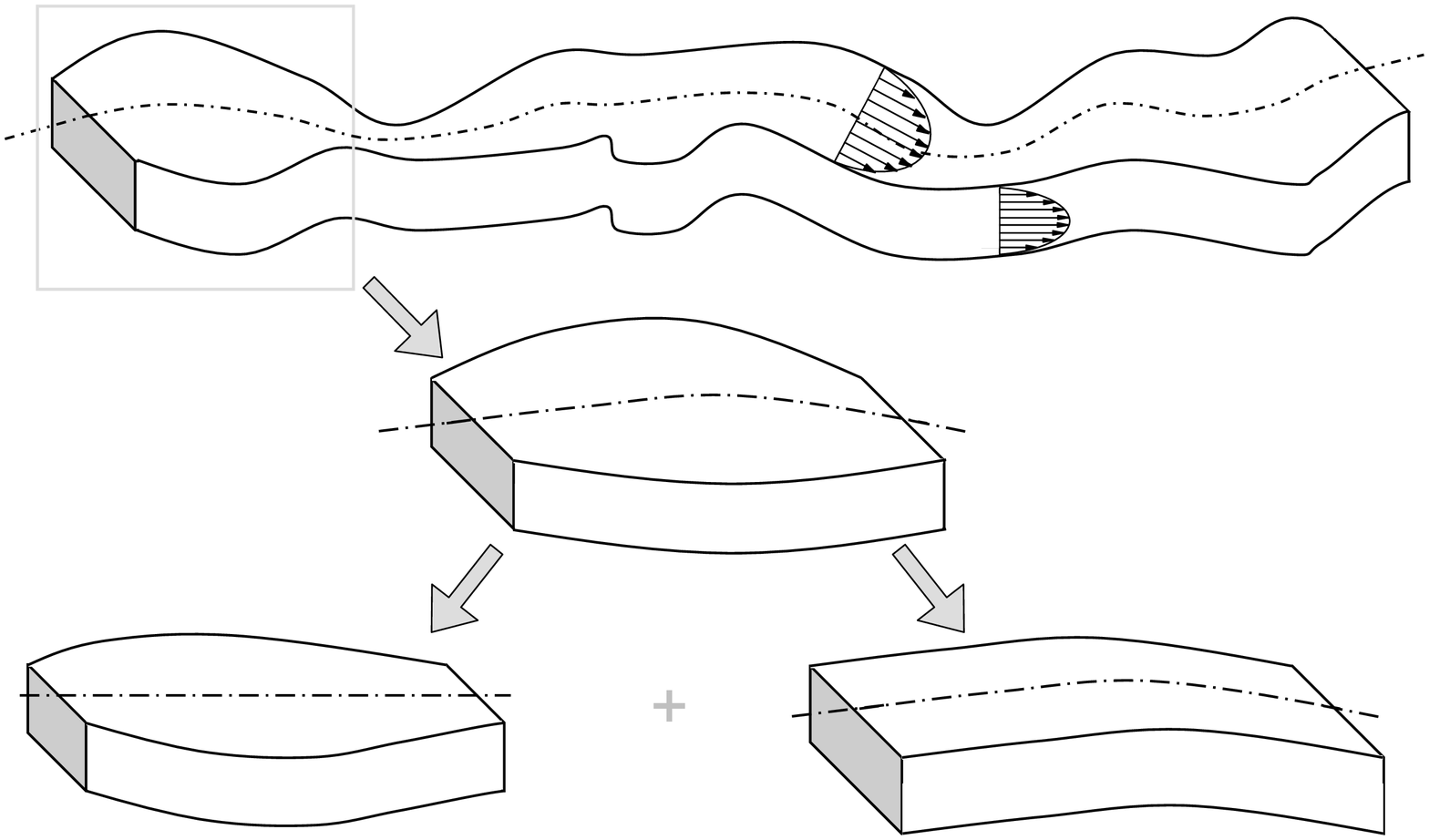}
\caption{Generic view of a microchannel constrained to remain
between two parallel planes. The design of channel has two degrees
of freedom: (1) the shape of its centerline and (2) the relative
width of the channel around this centerline.} \label{general}
\end{figure}

A variety of active (\cite{ultrasonic,Ho}) and passive
(\cite{Liu,Bessoth,swiss,Stroock,EOF}) methods have recently been
proposed to generate stirring flows for mixing purposes. All of
the passive designs that have been demonstrated to be effective
rely however on multilayer or non-planar structures in order to
generate three-dimensional flows. A single lithographic step
generates a single flat layer of structure with horizontal top and
bottom walls and (relatively) vertical side walls (see Figure
\ref{general}) (\cite{Kovacs}). More complicated structures, e.g.
a non-intersecting cross-over between two channels, require
multiple lithographic steps, with spatial alignment between each
layer of structure. Each added layer of structure increases the
difficulty of fabrication and complicates scaling down to smaller
devices.

Motivated by such practical considerations in this paper, we will
be interested in characterizing the potential for mixing of the
simplest planar geometrical configurations obtained in a single
fabrication step. Two such geometries are illustrated in Figure
\ref{general}. Suppose we fabricate a channel constrained between
two parallel planes of constant separation, can such a
configuration mix? In order to be able to give an answer to this
question that is robust to change in flow conditions, we will
assume zero Reynolds number flows in the channel. This assumption
also implies that our conclusions will remain valid in smaller
flow systems of the same design.

It is known that the steady-state velocity field in a straight
channel of constant rectangular cross section is unidirectional
(\cite{Batchelor}) and therefore cannot mix except by molecular
diffusion; similarly, the velocity field in a curved channel of
constant cross-section and constant curvature is unidirectional
(\cite{Rieger})\footnote{Note that both these results are in fact
valid independently of the cross-sectional shape of the channel as
long as the shape remains constant. Note also that time-periodic
flows in these geometries are also unidirectional and cannot be
used for low-Reynolds number mixing (see e.g. the review in
\cite{Ottino}).}. As a consequence, the simplest potential design
for a {\it steady-state} mixer for Stokes flow is that of a
channel with variations of shape, that include changes of both
curvature and cross-sectional dimensions in the streamwise
direction.

In order for flow in such a channel to potentially mix by
advection in the three dimensions of the channel, the velocity
field needs three non-zero components. While it is clear that
these variations in shape will lead to non-zero in-plane
components of the velocity, as would also be the case in a truly
two-dimensional channel, it is not obvious that the (third)
out-of-plane component will {\it always} be non-zero. We ask
therefore the following question: {\it under which circumstances
is the out-of-plane component of the velocity field always
non-zero?} And in this case, what is the expected magnitude of the
vertical flow?

The flows in a circular pipe of varying cross-section
(\cite{Manton}) or varying small curvature (\cite{Inaba}) have
been studied and three-dimensional flow is obtained at zero
Reynolds number. However, because the equation for the shape of a
circular pipe couples the two directions that are perpendicular to
its axis of symmetry, these results cannot be applied to the flow
in a planar geometry and a separate analysis has to be carried
out. Recently, Balsa \cite{Balsa} studied the secondary flow in a
Hele-Shaw cell in which a vertical cylinder is immersed, at
Reynolds number unity based on the cylinder length, and showed the
presence of streamwise vorticity in a boundary layer on the
cylinder surface; an earlier study by Thompson \cite{Thompson}
focused on viscous features.

The geometry of a generic microchannel constrained between two
parallel planes with fixed separation and with no obstacles is
illustrated in Figure \ref{general} (top). The shape of the
channel can be entirely described by two degrees of freedom: (1)
the shape of its centerline plane and (2) the local symmetric
width of the channel around this centerline. We will consider in
this paper the consequences of both and will treat each of them
separately for simplicity.

The paper is organized as follows. In section \ref{straight} we
consider the case of a straight channel with varying cross section
in the direction perpendicular to both the flow and the constraint
plane and in section \ref{curved} we consider the case of a curved
channel of constant cross section but varying curvature. In both
cases, under the assumption of an arbitrary but slowly varying
cross section and curvature respectively, we show that the
velocity component perpendicular to the constraint plane cannot be
zero unless cross-section and curvature are both constant, and
therefore the flow is fully three-dimensional in all other cases.
We apply these results in section \ref{application} where we
calculate the leading-order velocity field in the case of a
straight channel of varying cross section and illustrate the flow
patterns on a sinusoidally varying channel. We conclude in section
\ref{discussion} with a discussion of both the practical
advantages and limitations that these results imply for mixing
design. Appendices \ref{A1} and \ref{A2} present proofs for some
of the results used in sections \ref{straight} and \ref{curved}
respectively and Appendix \ref{asym} presents the calculation for
the leading-order velocity field in the case of a channel of
arbitrary shape.

\section{Three-dimensionality of the flow}
\subsection{Straight microchannel of varying cross section}
\begin{figure}
\centering
\includegraphics[width=.7\textwidth]{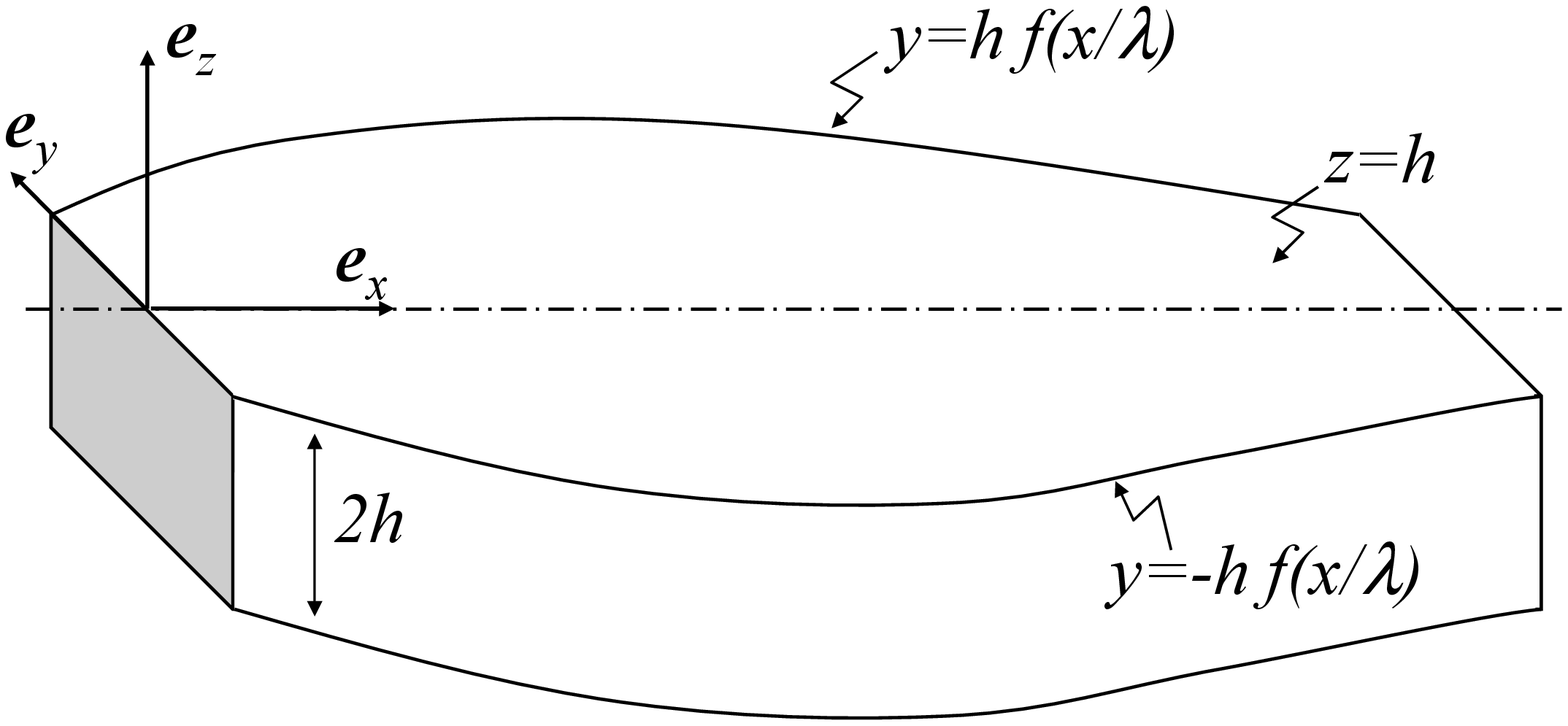}
\caption{Straight channel of slowly varying cross section in the $y$ direction.}
\label{fig1}
\end{figure}
In this section we consider the case of a straight microchannel of
varying cross section, as illustrated in Figure \ref{fig1}. A
pressure-driven flow takes place in the $x$ direction of a channel
of constant height $2h$ in the $z$-direction and varying width
$2hf(x/\lambda)$ in the $y$ direction, where $\lambda$ is the
axial length scale on which such variations occur. The equations
of motion and mass conservation for an incompressible Stokes flow
are written
\begin{equation} \nabla p=\mu
\nabla^2 {\bf u} ,\quad \nabla.{\bf u}=0 \label{stokes}
\end{equation}
with the no-slip boundary condition ${\bf u}(x,y=\pm
hf(x/\lambda),z)={\bf u}(x,y,z=\pm h)={\bf 0}$, on the four
bounding surfaces. The flow rate $Q$ is set by upstream conditions
and is constant
\begin{equation}
\int_{-h}^{h} \int_{-hf(x/\lambda)}^{+hf(x/\lambda)} ({\bf u}.{\bf
e}_x) \, \d y\,\d z
 =Q. \label{Q}
\end{equation}
We now make the assumption that the width of the channel is slowly
varying. If we define $\e=h/\lambda$, this assumption is
equivalent to assuming that $\e\ll 1$. Using the notations ${\bf
u}=(u,v,w)$ for the velocity field, we can now nondimensionalize
the previous set of equations \eqref{stokes}-\eqref{Q} by scaling
lengths, velocities and pressure by
\begin{equation}
(x,y,z)=(\lambda {\tilde x},h{\tilde y},h{\tilde z}),\quad\quad
(u,v,w)=\frac{Q}{h^2}({\tilde u},\e {\tilde v},\e {\tilde
w}),\quad\quad p=\frac{\lambda \mu Q}{h^4} {\tilde p}.
\label{scal}
\end{equation}
Dropping the tildes in the dimensionless variables for
convenience, the dimensionless Stokes equation is
\begin{equation}\label{stokes_cartesian}
\nabla p   = \,\,\,\,\,\, \left(\epsilon^2\frac{\p^2}{\p x^2}+
\frac{\p^2}{\p y^2}+ \frac{\p^2}{\p z^2} \right) \Big[u,\epsilon^2
v, \epsilon^2 w\Big]
\end{equation}
and the dimensionless continuity equation is
\begin{equation}\label{cont2}
\frac{\p u}{\p x} + \frac{\p v}{\p y} +\frac{\p w}{\p z} = 0.
\end{equation}
We look for a regular perturbation expansion for both the velocity
and pressure fields under the form
\begin{equation}\label{exp}
(u,v,w,p)  = (u_0, v_0, w_0,p_0) +\e^2(u_2,v_2,w_2,p_2)+ {\cal
O}(\e^4)
\end{equation}
which is usual in lubrication theory (see e.g. \cite{ghosal}). The
leading-order ${\cal O}(\e^0)$ term of Stokes equation
\eqref{stokes} is given by
\begin{equation}\label{first}
\frac{\p p_0}{\p x}  =  \left(\frac{\p^2}{\p y^2}+\frac{\p^2}{\p
z^2} \right)u_0 ,\quad
 \frac{\p p_0}{\p y}  =   \frac{\p p_0}{\p z}= 0
\end{equation}
subject to the no-slip boundary condition $u_0(x,y=\pm
f(x),z)=u_0(x,y,z=\pm 1)=0$, and constant flow rate
\begin{equation}
4 \int_{0}^{1} \int_{0}^{f(x)} u_0 \, \d y\, \d z =1. \label{flow}
\end{equation}
The axial velocity $u_0$ is then easily calculated by separation
of variables (\cite{Batchelor})
\begin{equation}u_0(x,y,z)=\frac{1}{2}\frac{\d
p_0}{\d x} \left\{ (z^2-1) + \sum_{n\geq
0}\frac{4(-1)^{n}}{k_n^3}\left(\frac{\cosh k_ny}{\cosh
k_nf(x)}\right)\cos k_nz \right\} \label{u0}
\end{equation}
with $k_n=(n+\frac{1}{2})\pi$. This solution is simply the axial
Poiseuille velocity in a straight channel of constant cross
section evaluated at each location along the channel. The
leading-order axial pressure gradient is then given by the flow
rate condition \eqref{flow} which leads to
\begin{equation} \frac{\d p_0}{\d x} =\frac{3}{4f(x)}\left\{
\frac{6}{f(x)}\sum_{n\geq 0}\frac{\tanh (k_nf(x))}{k_n^5} -1
\right\}^{-1}\cdot \label{gradp}
\end{equation} \label{straight}
Note that \eqref{stokes_cartesian} and \eqref{cont2} show that at
next order in $\e^2$, the leading-order out of plane velocity
field $(v_0,w_0)$ satisfies a two-dimensional Stokes equation with
an effective distribution of mass sources and sinks given by $-\d
u_0 / \d x $. Let us now assume these sources and sinks lead to a
planar velocity field is planar in the sense that the component of
the velocity perpendicular to the constraint plane is zero,
$w_0=0$. In this case, the continuity equation from \eqref{stokes}
is written
\begin{equation}
\frac{\p u_0}{\p x} +
\frac{\p v_0}{\p y}   =  0, \label{cont}
\end{equation}
and allows us to solve explicitly for the $y$-component of the
velocity $v_0$. Using the fact that $u_0$ satisfies the no-slip
condition on the walls of the channels, it is straightforward to
obtain that $v_0$ is given by
\begin{equation}
v_0(x,y,z)= \frac{\p}{\p x} \int_{-f(x)}^y u_0(x,y',z) \,\d y'.
\label{v0}
\end{equation}
The solution \eqref{v0} satisfies the no-slip  conditions for
$v_0$ at $z=\pm 1$ and $y=-f(x)$. If it also satisfies the
remaining no-slip condition at $y=f(x)$ then the leading-order
solution would be entirely characterized and the flow would be
planar at leading-order. The condition at $y=f(x)$ will however be
satisfied if and only if
\begin{equation}
\frac{\p}{\p
x} \int_{-f(x)}^{f(x)} u_0(x,y',z) \,\d y'=0
\label{cond}
\end{equation}
for all values of $x$ and $z$ in the channel. Using
the solution \eqref{u0}, \eqref{cond} can be integrated once to
get
\begin{equation}
\frac{\d p_0}{\d x} \left\{ \sum_{n\geq
0}\frac{(-1)^{n}}{k_n^4}(\tanh (k_nf(x))-k_n)\cos k_nz \right\}
=\Phi(z). \label{condition}
\end{equation}
In order for equation \eqref{condition} to be satisfied for all
$|z|\leq 1$ and $x\geq 0$, it is then necessary that for all
$n\geq 0$,
\begin{equation}
 \frac{\d p_0}{\d x} (\tanh (k_nf(x))-k_n)=\delta_n
\label{constant_straight}
\end{equation} where the $\{\delta_n\}$ are constants independent of
$x$. As is shown in Appendix \ref{A1}, this can only be true if
$f(x)$ is constant, {\it i.e.} if the channel cross section is
constant. As a consequence, the vertical component of the velocity
field $w_0$ cannot be zero unless the cross section of the channel
is constant, in which case $v_0=w_0=0$.  When this is not the case
and the cross section is changing along the channel, then the
two-dimensional solution \eqref{u0}-\eqref{v0} is inconsistent and
the velocity field is fully three-dimensional at this order.


\subsection{Channel of constant cross section with varying curvature}
\label{curved} 
\begin{figure}
\centering
\includegraphics[width=.7\textwidth]{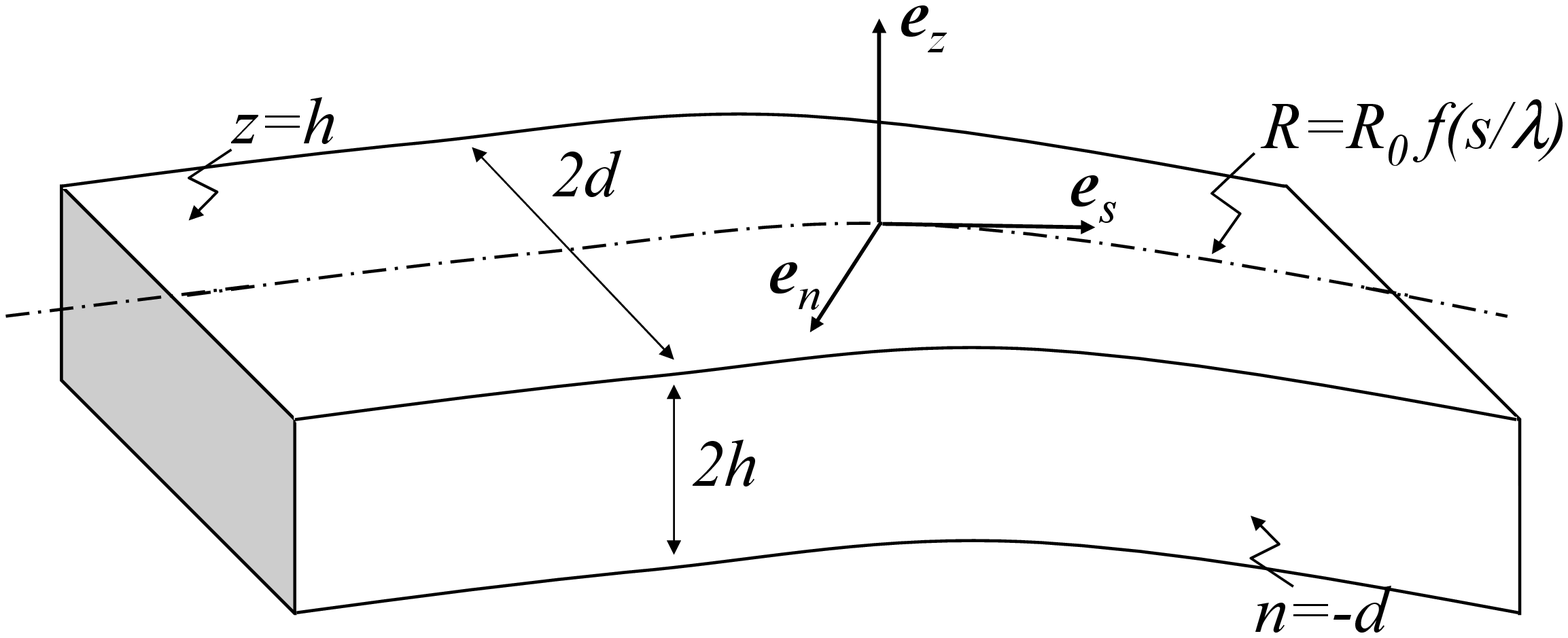}
\caption{Curved microchannel of constant cross section and slowly
varying planar curvature.} \label{fig2}
\end{figure}

We now proceed in the same manner as in section
\ref{straight} for the case of the channel of constant cross
section but varying curvature, as illustrated in Figure
\ref{fig2}. A pressure-driven flow takes place in the axial
direction, denoted as $s$, of a channel of constant height $2h$ in
the $z$-direction and constant width $2d$ in the third direction,
denoted as $n$ for ``normal''. The centerline of the channel is
not straight but curved with local radius of curvature
$R(s)=R_0f(s/\lambda)$ in the orthogonal $({\bf e}_n,{\bf
e}_s,{\bf e}_z)$ frame, where $\lambda$ is the typical length
scale along the channel on which this local curvature changes. In
this geometry, and using the notations ${\bf u}=(u,v,w)$ for the
velocity field, it is possible after some algebra to write the
dimensional Stokes equation \eqref{stokes} under the form
\begin{subeqnarray} \label{stokes_curved}
\frac{1}{\mu}\frac{R(s)}{R(s)+n}\frac{\p p}{\p s}&= & \frac{\p^2u
}{\p z^2} + \left(\frac{R(s)}{R(s)+n}\right)^2 \frac{\p^2 u}{\p
s^2}
+ \frac{\p }{\p n}\left(\frac{1}{R(s)+n}\frac{\p}{\p n}(R(s)+n)u \right)\nonumber  \\
& &
+\frac{2R(s)}{(R(s)+n)^2} \frac{\p v}{\p s}+\frac{R(s)}{(R(s)+n)^2} \frac{\p v}{\p n}\frac{\d R }{\d s},\\
\frac{1}{\mu}\frac{\p p}{\p n}&=& \frac{\p^2v }{\p z^2} +
\left(\frac{R(s)}{R(s)+n}\right)^2 \frac{\p^2 v}{\p s^2}
+ \frac{\p }{\p n}\left(\frac{1}{R(s)+n}\frac{\p}{\p n}(R(s)+n)v \right)\nonumber \\
 & &
-\frac{2R(s)}{(R(s)+n)^2} \frac{\p u}{\p s}
+\frac{R(s)}{(R(s)+n)^3} \frac{\d R }{\d s} \left(n\frac{\p v}{\p s}-u\right),\\
\frac{1}{\mu}\frac{\p p}{\p z}&=&\frac{\p^2w }{\p z^2} +
\left(\frac{R(s)}{R(s)+n}\right)^2 \frac{\p^2 w}{\p s^2} +
\frac{1}{R(s)+n}\frac{\p }{\p n}\left((R(s)+n)\frac{\p w }{\p
n}\right) \nonumber \\ && +\frac{nR(s)}{(R(s)+n)^3}\frac{\d R }{\d
s} \frac{\p w}{\p s},
\end{subeqnarray}
and the continuity equation is written
\begin{equation}
\left(\frac{R(s)}{R(s)+n}\right)\frac{\p u}{\p s} +
\frac{1}{R(s)+n}\frac{\p }{\p n}\left((R(s)+n) v \right) +\frac{\p
w}{\p z} =0. \label{cont_curved}
\end{equation}
The two sets of equations \eqref{stokes_curved} and
\eqref{cont_curved} are associated with the no-slip boundary
condition on the walls of the channel ${\bf u}(s,n=\pm d,z)={\bf
u}(s,n,z=\pm h)=0$, as well as with the condition of constant flow
rate along the channel
\begin{equation} \int_{-h}^{h}
\int_{-d}^{d} ({\bf u}.{\bf e}_s) \, \d z\, \d n =Q.
\label{Qcurved}
\end{equation}
As in section \ref{straight}, we now assume a slowly varying
curvature, {\it i.e.} we assume that both $h/\lambda\ll 1$ and
$d/\lambda\ll 1$. Equations \eqref{stokes_curved} -
\eqref{Qcurved} can be nondimensionalized by scaling lengths,
velocities and pressure by
\begin{equation}
(s,n,z)\sim (\lambda {\tilde s}, d{\tilde n},h{\tilde
z}),\quad\quad (u,v,w)=\frac{Q}{hd}({\tilde
u},\frac{\e}{\alpha}{\tilde v},\e {\tilde w}),\quad\quad
p=\frac{\lambda \mu Q}{h^3d} {\tilde p}
\end{equation}
where we denoted the aspect ratio $\alpha=h/d={\cal O}(1)$.
Defining $\beta=d/R_0$ and $\e=h/\lambda\ll 1$, and dropping the
tildes in the dimensionless variables for convenience, the
dimensionless Stokes equation is
\begin{subeqnarray} \label{stokes_curved_2}
\frac{f(s)}{f(s)+\beta n}\frac{\p p}{\p s}&= & \frac{\p^2u }{\p
z^2} + \epsilon^2\left(\frac{ f(s)}{f(s)+\beta n}\right)^2
\frac{\p^2 u}{\p s^2}
+ \frac{\p }{\p n}\left(\frac{\alpha^2}{f(s)+\beta n}\frac{\p}{\p n}(f(s)+\beta n)u \right)\nonumber \\
 & &
+\frac{2\epsilon^2\beta f(s)}{(f(s)+\beta n)^2} \frac{\p v}{\p s}+\frac{\epsilon^2 f(s)f'(s)}{(f(s)+\beta n)^2} \frac{\p v}{\p n},\\
\frac{\p p}{\p n}&=& \frac{\epsilon^2}{\alpha^2}\frac{\p^2v }{\p
z^2} + \frac{\epsilon^4}{\alpha^2}\left(\frac{f(s)}{f(s)+\beta
n}\right)^2 \frac{\p^2 v}{\p s^2} + \epsilon^2\frac{\p }{\p
n}\left(\frac{1}{f(s)+\beta n}\frac{\p}{\p n}(f(s)+\beta n)v
\right)
\nonumber \\
 & & -\frac{2\epsilon^2\beta f(s)}{(f(s)+\beta n)^2}
\frac{\p u}{\p s}
+\frac{\epsilon^3\beta f(s)f'(s)}{\alpha(f(s)+\beta n)^3}  \left(n\frac{\epsilon}{\alpha}\frac{\p v}{\p s}-u\right),\\
\frac{\p p}{\p z}&=&\epsilon^2\frac{\p^2w }{\p z^2} +
\epsilon^4\left(\frac{f(s)}{f(s)+\beta n}\right)^2 \frac{\p^2
w}{\p s^2} + \frac{\epsilon^2\alpha^2}{f(s)+\beta n}\frac{\p }{\p
n}\left((f(s)+\beta n)\frac{\p w }{\p n}\right)\nonumber \\ &&
+\frac{n\epsilon^4\beta f(s)f'(s)}{(f(s)+\beta n)^3} \frac{\p
w}{\p s},
\end{subeqnarray}
and the dimensionless continuity equation
\begin{equation}
\frac{f(s)}{f(s)+\beta n} \frac{\p u}{\p s} +\frac{1}{f(s)+\beta
n} \frac{\p }{\p n}(f(s)+\beta n) v+\frac{\p w}{\p z} =0.
\end{equation}
Note that $\beta$ is not necessary small in actual MEMS
applications \cite{isma}. We then look for a regular perturbation
expansion for both the dimensionless velocity and pressure fields
under the form
\begin{equation}\label{exp2}
(u,v,w,p)  = (u_0, v_0,w_0,p_0) +\e^2(u_2,v_2,w_2,p_2)+ {\cal
O}(\e^4).
\end{equation}
The leading-order ${\cal O}(\e^0)$ of the Stokes equation
\eqref{stokes_curved}  is
\begin{subeqnarray}\label{u0_curved}
\frac{f(s)}{f(s)+\beta n} \frac{\p p_0}{\p s}  & = & \frac{\p^2u_0
}{\p z^2} + \frac{\p }{\p n}\left(\frac{\alpha^2 }{f(s)+\beta
n}\frac{\p}{\p n}(f(s)+\beta n)u_0 \right)\\
\frac{\p p_0}{\p n} & = &\frac{\p p_0}{\p z}  =0
\end{subeqnarray}
together with the no-slip boundary condition $u_0(s,n=\pm
1,z)=u_0(s,n,z=\pm 1)=0$, and constant flow rate
\begin{equation}
\int_{-1}^{1}
\int_{-1}^{1}
u_0
\,
\d z\,
\d n
=1.
\label{Qcurved_adim}
\end{equation}
Using separation of variables, it is possible to solve for the
axial velocity component in equations \eqref{u0_curved} -
\eqref{Qcurved_adim}, similarly to what was done by Rieger \&
S\u{e}st\'ak (1973) for the case of a curved rectangular channel
of constant curvature. We obtain
\begin{equation}
u_0(s,n,z)=f(s)\frac{\d p_0}{\d s}
\left\{\frac{z^2-1}{2(f(s)+\beta n)} +\sum_{n\geq 0}
U_n\left(\frac{k_n(f(s)+\beta n)}{\alpha\beta}\right) \cos k_nz
\right\}, \label{u0curved}
\end{equation}
where the set of functions $U_n$ are defined by
\begin{equation}
U_n(\eta)= E_n (s)K_1\left(\eta\right)
+F_n(s)I_1\left(\eta\right),
\end{equation}
with $I_1$ and $K_1$ as the order-one modified Bessel functions of
the first and second kind respectively, and with
\begin{subeqnarray}
\label{EnFn} \slabel{En} E_n(s) & = & \left\{
(f(s)-\beta)I_1(k_n^-(s))- (f(s)+\beta)I_1(k_n^+(s)) \right\}
G_n(s),\\
\slabel{Fn} F_n(s) & = & \left\{ (f(s)+\beta)K_1(k_n^+(s))-
(f(s)-\beta)K_1(k_n^-(s)) \right\} G_n(s)
\end{subeqnarray}
where
\begin{equation}
\label{Gn} G_n(s)=\frac{2(-1)^n}{k_n^3(f(s)^2-\beta^2)} \left\{
I_1(k_n^-(s)) K_1(k_n^+(s)) - I_1(k_n^+(s)) K_1(k_n^-(s)) \right\}
^{-1},
\end{equation}
and
\begin{equation}
k_n^{\pm}(s)=\frac{k_n(f(s)\pm\beta )}{\alpha\beta} \cdot
\end{equation}
Using the identities $K_0'=-K_1$ and $I_0'=-I_1$ and the flow rate
condition \eqref{Qcurved_adim} we obtain the pressure gradient
\begin{equation}
\frac{\d p_0}{\d s}=\frac{1}{2f(s)} \left\{ \alpha\sum_{n\geq
0}(-1)^n\frac{H_n(s)}{k_n^2} -\frac{1}{3\beta}\ln
\left(\frac{f(s)+\beta}{f(s)-\beta}\right) \right\}^{-1},
\label{pressure_curved}
\end{equation}
with
\begin{equation}
H_n(s)=E_n(s) \left\{ K_0(k_n^-(s))- K_0(k_n^+(s)) \right\} +F_n
\left\{ I_0(k_n^+(s))- I_0(k_n^-(s)) \right\}\cdot \label{Hn}
\end{equation}

As in section \ref{straight}, let us now make the assumption that
the flow is planar, {\it i.e.} that the leading-order vertical
component of the velocity field is zero, $w_0=0$. In this case the
continuity equation \eqref{cont_curved} becomes
\begin{equation}
\frac{\p u_0}{\p s} +\frac{1}{f(s)} \frac{\p }{\p n}(f(s)+\beta n)
v_0 =0.
\end{equation}
which can be used to solved exactly for $v_0$
\begin{equation}
v_0(s,n,z)=-\frac{f(s)}{f(s)+\beta n}\frac{\p}{\p s} \left\{
\int_{-1}^nu_0(t,n',z)\,\d n' \right\}\cdot \label{v0_curved}
\end{equation}
The solution \eqref{v0_curved} satisfies the no-slip boundary
condition at $z=\pm 1$ and $n=-1$; if the condition at $n=1$ was
also satisfied, the leading-order velocity field would be
two-dimensional at leading-order, $w_0=0$. The no-slip boundary
condition evaluated at $n=1$ will however be satisfied if and only
if
\begin{equation}
\frac{\p}{\p s} \left\{ \int_{-1}^1 u_0(t,n',z)\,\d n' \right\}=0,
\label{noslip_curved}
\end{equation}
for all values of $s$ and $z$. Using the solution for the axial
velocity \eqref{u0curved}, \eqref{noslip_curved} can be integrated
once to obtain \be f(s) \frac{\d p_0}{\d s}\left\{ \sum_{n\geq
0}\cos k_n z \left( \frac{\alpha H_n(s)}{k_n}-\frac{2(-1)^n}{\beta
k_n^3\ln\left(\frac{f(s)+\beta}{f(s)-\beta}\right)} \right)
\right\} =\Psi(z), \label{C2}
\end{equation}
where $H_n$ is defined in \eqref{Hn}. In order for \eqref{C2} to
be satisfied for all $|z|\leq 1$ and $s\geq 0$, it is necessary
that all for all $n\geq 0$,
\begin{equation}
f(s)\frac{\d p_0}{\d s} \left\{ \frac{\alpha
H_n(s)}{k_n}-\frac{2(-1)^n}{\beta
k_n^3\ln\left(\frac{f(s)+\beta}{f(s)-\beta}\right)}
\right\}=\gamma_n \label{constant_curved}
\end{equation}
where the $\{\gamma_n\}$ are constants independent of $s$. As is
shown in Appendix \ref{A2}, \eqref{constant_curved} can be
satisfied if and only if $f(s)$ is constant, {\it i.e.} if the
curvature of the channel is constant, in which case $v_0=w_0=0$.
When this is not the case and the curvature is changing along the
channel, then the two-dimensional solution
\eqref{u0curved}-\eqref{v0_curved} is inconsistent and the
velocity field is fully three-dimensional at this order.

\section{Illustration of the three-dimensional flows}
\label{application}

We have demonstrated in the previous two sections that flows in
channels constrained to remain in a layer of constant thickness
are in general three-dimensional, {\it i.e.} they possess a
non-zero component of the velocity perpendicular to the constraint
plane. We illustrate these results in this section for the case
studied in section \ref{straight} of a straight channel of varying
width. We calculate the three components of the leading-order
velocity field $(u_0,v_0,w_0)$ and illustrate the flow patterns in
a sinusoidally varying channel; the calculation for the general
case of an asymmetric channel is more intricate and we present it
in Appendix \ref{asym} for the interested reader.

\subsection{Governing equations}
Because the velocity field $(u_0,v_0,w_0)$ is three-dimensional,
the continuity equation in \eqref{stokes} becomes
\begin{equation} \frac{\p u_0}{\p
x} + \frac{\p v_0}{\p y} + \frac{\p w_0}{\p z} = 0 \label{cont3D},
\end{equation}
where $u_0$ is still given by equation \eqref{u0}. Under Stokes
flow conditions, the pressure is harmonic $\nabla^2 p=0$, and
therefore the velocity field always satisfies the biharmonic
equation $\nabla^4 {\bf u}=0$. Consequently, within the
lubrication approximation \eqref{scal} and \eqref{exp}, the $y$
and $z$ component of the dimensionless velocity field satisfy
\begin{subeqnarray} \slabel{biv} \np^4 v_0 & = & 0, {\rm\quad where\quad}
\np^2 \equiv \left(\frac{\p^2}{\p y^2}+\frac{\p^2}{\p
  z^2}\right)\\
\slabel{biw} \np^4 w_0 & = & 0. \label{bi}
\end{subeqnarray}
Similarly, the vorticity is harmonic $\nabla^2 {\boldsymbol{
\omega}}=0$, so that under the lubrication approximation, its
leading-order axial component $\omega_0=\frac{\p w_0}{\p
y}-\frac{\p v_0}{\p z}$ satisfies
\begin{equation}
\np^2 \omega_0=0 \label{vort}.
\end{equation}
It is necessary to solve the set of equations \eqref{cont3D},
\eqref{bi} and \eqref{vort} along with the no-slip boundary
conditions in order to obtain the final solution for $(v_0,w_0)$.

\subsection{Subset of equations}
\label{subset} Let us now show that it is sufficient to solve
equations \eqref{cont3D} and \eqref{biv} to obtain \eqref{biw} and
\eqref{vort}. Let us suppose \eqref{cont3D} and \eqref{biv} are
satisfied. Evaluating the bi-harmonic $\np^2$ of the continuity
equation \eqref{cont3D} leads to
\begin{equation}
\frac{\p }{\p z}\np^4 w_0=0 \rightarrow \np^4 w_0 = \Gamma(x,y).
\end{equation}
Because of the symmetries in the configuration  illustrated in
Figure \ref{general}, the solution of Stokes equation has to
satisfy $w_0(x,y,-z)=-w_0(x,y,z)$ (and also
$v_0(x,y,-z)=v_0(x,y,z)$). Consequently $\np^4 w_0$ is also odd
with respect to $z$ and necessarily $\Gamma(x,y)=0$, so that
equation \eqref{biw} is satisfied. In the same fashion, it is
straightforward from \eqref{cont3D} to obtain \be \frac{\p
\omega_0}{\p z} =0 \rightarrow \omega_0 =\Lambda(x,y).
\end{equation}
Using the fact that $w_0$ and $v_0$ are respectively  odd and even
with respect to $z$, it is clear that $\omega_0 $ is odd with
respect to $z$ so that $\Lambda(x,y)=0$. The result of equation
\eqref{vort} is therefore recovered. As a consequence, it is
sufficient to solve equations \eqref{cont3D} and \eqref{biv} to
obtain the complete solution for the leading-order velocity field.

\subsection{Velocity field calculation}
\label{solution} To obtain the leading-order solution for the
three-dimensional velocity field in the channel, we solve
equations \eqref{biv} and \eqref{cont3D}, along with the no-slip
boundary conditions for $v_0$ and $w_0$ and with the axial
velocity $u_0$ given by \eqref{u0}. In order to do so, we use the
technique introduced more than a century ago by Lam\'e
(\cite{Lame}) to solve planar elasticity problem where biharmonic
equations arise (see also the general discussion in
\cite{Meleshko}). Here we effectively demonstrate that these ideas
also apply as well to slowly varying flows. Using the following
symmetries in the velocity field
\begin{subeqnarray} \slabel{symv}
v_0(x,y,-z)  &= &  v_0(x,y,z),\quad \,\,\,v_0(x,-y,z)  = - v_0(x,y,z),\\
w_0(x,-y,z)  &= &  w_0(x,y,z),\quad w_0(x,y,-z)  = - w_0(x,y,z),
\label{sym}
\end{subeqnarray}
we look for a solution of \eqref{biv} under the form of a double
Fourier series in $y$ and $z$
\begin{equation}
v_0(x,y,z)=\sum_{n\geq 0} A_n(x,y) \cos k_nz +\sum_{m> 0} B_m(x,z)
\sin\left( \frac{\ell_m y}{f(x)}\right), \label{v03D}
\end{equation}
with $\ell_m=m\pi$ and $k_n=(n+1/2)\pi$. In order for \eqref{biv}
and \eqref{symv} to be satisfied, the functions $A_n$ and $B_m$
are given by\footnote{Note that in \eqref{An} we use the indices
$n$ and $m$ twice but do not imply this is an implicit summation
notation.}
\begin{equation} A_n(x,y)=a_n(x)P_n(x,y),\quad
B_m(x,z)=b_m(x)Q_m(x,z) \label{An}
\end{equation}
with
\begin{subeqnarray}
P_n(x,y) & = & f(x)\,{\sinh (k_n y)} -
{y}\cosh (k_n y)\tanh (k_n f(x))\\
Q_m(x,z) & = & {\tanh \left( \frac{\ell_m }{f(x)}\right) }\cosh
\left( \frac{\ell_m z}{f(x)}\right) - z \,{\sinh \left(
\frac{\ell_m z}{f(x)}\right)}
\end{subeqnarray}
and where both ($a_n(x)$) and ($b_m(x)$) are unknown functions to
be determined. With the axial solution \eqref{u0} written
\begin{equation}\label{un}
u_0(x,y,z)=\sum_{n\geq 0} {\cal U}_n(x,y) \cos k_nz,\quad {\cal
U}_n(x,y)=\frac{2(-1)^n}{k_n^3}\frac{\d p_0}{\d
x}\left(\frac{\cosh (k_n y)}{\cosh (k_nf(x))}-1\right)
\end{equation}
and with \eqref{v03D}, integration of the continuity equation
\eqref{cont3D} leads to the third component of the velocity field
\be w_0(x,y,z)=-\sum_{n\geq 0}\frac{1}{k_n}\left( \frac{\p {\cal
U}_n}{\p x} + a_n(x)\frac{\p P_n}{\p y} \right)\sin k_nz -\sum_{m
> 0}\frac{\ell_m b_m(x)T_m(x,z) }{f(x)}\cos \left(\frac{\ell_m
y}{f(x)}\right) \label{w03D}
\end{equation}
with
\begin{equation}
T_m(x,z)= \left(\frac{f(x)}{\ell_m}\tanh \left(\frac{\ell_m
}{f(x)}\right)+
\frac{f(x)^2}{\ell_m^2}\right)\sinh\left(\frac{\ell_m
z}{f(x)}\right)-\frac{f(x)}{{\ell_m}}z\cosh \left(\frac{\ell_m
z}{f(x)}\right)\cdot
\end{equation}
The sets of unknown functions ($a_n$) and ($b_m$) are finally
determined by enforcing the no-slip boundary condition for the
solution \eqref{w03D} at both $z=\pm 1$ and $y=\pm f(x)$. As is
usually the case for problems involving biharmonic equations (e.g.
\cite{Meleshko}), the final result involves an infinite system of
linear algebraic equations given by
\begin{equation}\left\{
\begin{array}{ccc}
a_n(x) & = & \displaystyle\overline{a_n}(x)+\sum_{m> 0}A_{nm}(x) b_m(x)\,\,\,\,(n\geq 0),\\
b_m(x) & = & \displaystyle\overline{b_m}(x)+\sum_{n\geq 0}
B_{mn}(x)  a_n(x) \,\,\,\,(m > 0), \label{dual}
\end{array}\right.
\end{equation}
with
\begin{subeqnarray}\label{def}
\overline{a_n}(x)&=& -\frac{\p {\cal U}_n}{\p x}(x,f(x)) \left\{
\frac{\p P_n}{\p y}(x,f(x)) \right\}^{-1}, \\
A_{nm}(x) & = & \frac{2(-1)^{m+1}\ell_mk_n}{f(x)} \left\{ \frac{\p
P_n}{\p y}(x,f(x)) \right\}^{-1} \int_{0}^{1} T_m(x,z)\sin k_n z
\,\d z
,\\
 \overline{b_m}(x)& = & \frac{2}{\ell_mT_m(x,1)}\sum_{n\geq 0}\frac{(-1)^{n+1}}{k_n}\int_{0}^f
\frac{\p {\cal U}_n}{\p x}(x,y)  \cos \left(\frac{\ell_m
y}{f(x)}\right)\, \d y \slabel{bmbar}
,\\
B_{mn}(x) & = & \frac{2(-1)^{n+1}}{k_n\ell_mT_m(x,1)}\int_{0}^f
\frac{\p P_n}{\p y}(x,y) \cos \left(\frac{\ell_m y}{f(x)}\right)
\, \d y.
\end{subeqnarray}
Note that, at a given position $x$, both the ($a_n(x)$) and
($b_n(x)$) are entirely determined by the instantaneous values of
$f(x)$ and $f'(x)$; each subsequent order in the long-wavelength
expansion \eqref{exp} will bring an additional  dependance on a
higher derivative of $f(x)$.

\subsection{Further calculations}
\subsubsection{Axial vorticity}

Given the sets of $\{a_n\}$ and $\{b_n\}$, we can evaluate the
axial component of the vorticity $\omega_0=\frac{\p w_0}{\p
y}-\frac{\p v_0}{\p z}$:
\begin{subeqnarray}\label{omega03D}
\omega_0 (x,y,z) &=&-\sum_{n\geq 0}\frac{1}{k_n}\left(\frac{\p^2
{\cal U}_n}{\p x\p y}+a_n(x)\frac{\p^2 P_n}{\p
y^2}-k_n^2a_n(x)P_n(x,y) \right) \sin
k_n z \\
&&+\sum_{m>0}\left( \frac{\ell_m^2 b_m(x)T_m(x,z) }{f(x)^2}-b_m(x)
\frac{\p Q_m}{\p z}\right) \sin\left( \frac{\ell_m
y}{f(x)}\right)\cdot
\end{subeqnarray}

\subsubsection{Quadrant-averaged velocities}

The quadrant-averaged velocities can also be evaluated, for
example in the quadrant ($y>0,\,z>0$). The flow rate condition
\eqref{flow} leads to a constant average axial velocity,
$<u_0>(x)={1}/{4}$. Integration of \eqref{v03D} and \eqref{w03D}
across the quadrant leads to
\begin{equation}\label{}
<v_0>(x)=\sum_{n\geq 0}\frac{(-1)^na_n}{k_n
f(x)}\int_0^{f(x)}P_n(x,y) \d y
 + \sum_{m>0}\frac{b_mf(x)(1+(-1)^{m+1})}{\ell_m}\int_0^1 Q_m(x,z)
 \d z,
\end{equation}
and
\begin{equation}\label{}
<w_0>(x)=\sum_{n\geq 0}\frac{2(-1)^{n+1}}{k_n^6f(x)}\frac{\d}{\d
x} \left[\frac{\d p_0}{\d x}\left(\tanh (k_n f(x)) -k_nf(x)
\right) \right].
\end{equation}

\subsection{Case of a sinusoidally varying wall}
\label{illustration}

We chose to illustrate the flow patterns in the case where the
wall shape is described by the dimensionless function $f(x)=1+0.7
\sin x$; recall that the actual dimensional wall shape is
 described by $f(\epsilon x)$ where $\epsilon = h/\lambda$. The infinite system of
linear equations \eqref{dual} was solved numerically by truncating
it at finite values of $n$ and $m$. The integrals in \eqref{def}
 only involve linear and trigonometric functions and are
evaluated exactly. Note that apart from the system \eqref{dual},
summations are also involved in equations \eqref{gradp} for the
pressure gradient and \eqref{bmbar} for $ \overline{b_m}(x)$, and
they also require numerical truncations.

For each case, numerical results were obtained, the truncation was
refined and the results were found to converge quickly to a final
solution. The truncations at $n=50$ in equations \eqref{gradp} and
\eqref{bmbar} were found to be suitable to obtain the final
solution. Further, a truncation at $n=m=20$ in the infinite set of
linear equations equation \eqref{dual} was also found to be
appropriate to resolve the flow fields, with results essentially
unchanged for higher truncation numbers.

Such techniques allow us to obtain everywhere in space the three
leading-order velocity components and therefore, with a simple
time advancement scheme, to follow the motion of individual fluid
elements and obtain streamlines.

\begin{figure}
\centerline{\includegraphics[width=.97\textwidth]{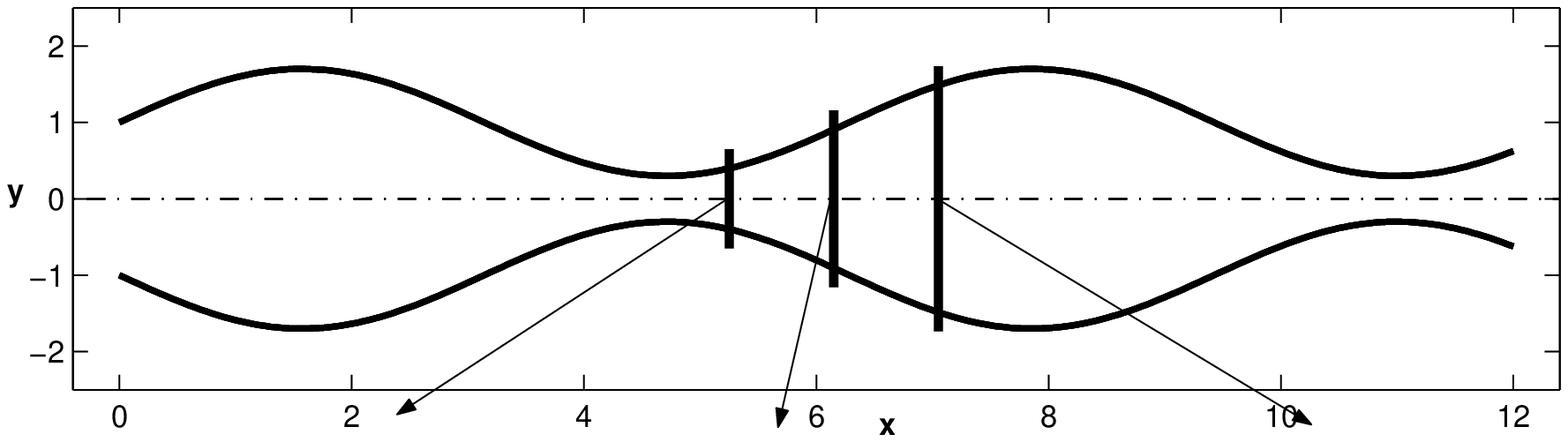}\,\,\,\,\,\quad\quad\quad}
\centerline{
\includegraphics[height=.2\textwidth]{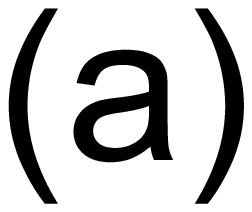}
\includegraphics[width=.26\textwidth]{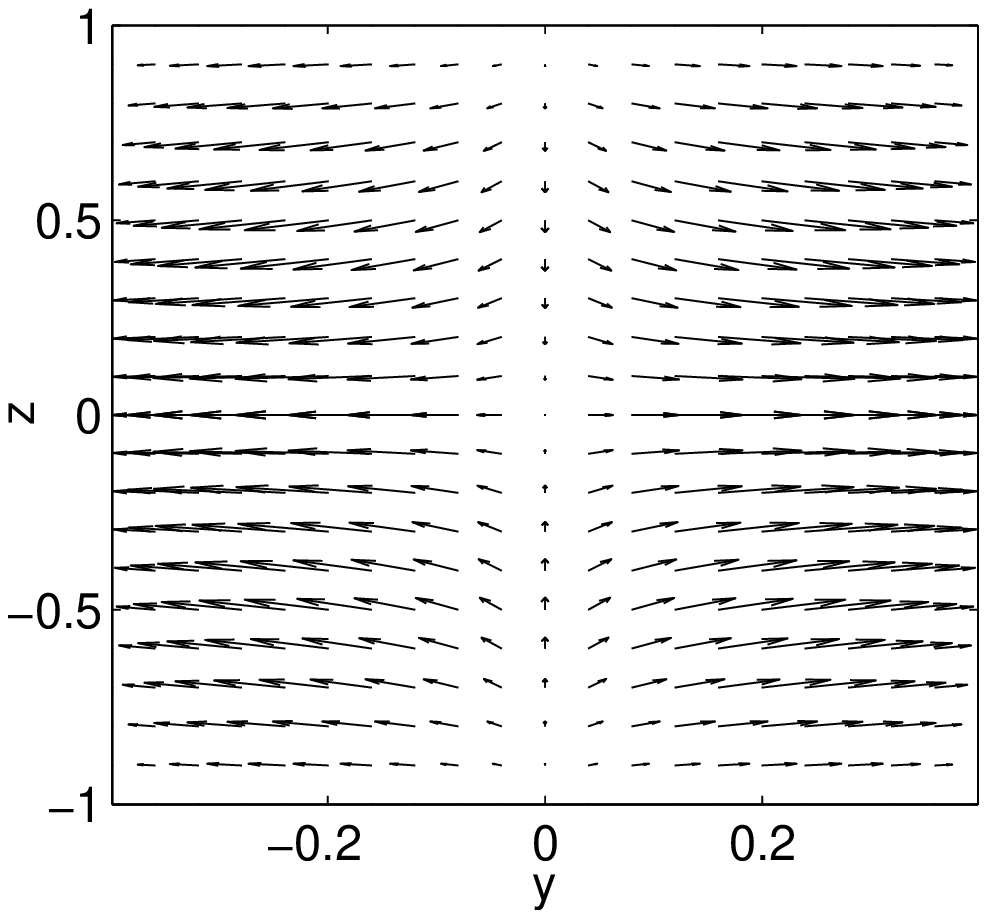}\,\,\,\,\,
\includegraphics[width=.26\textwidth]{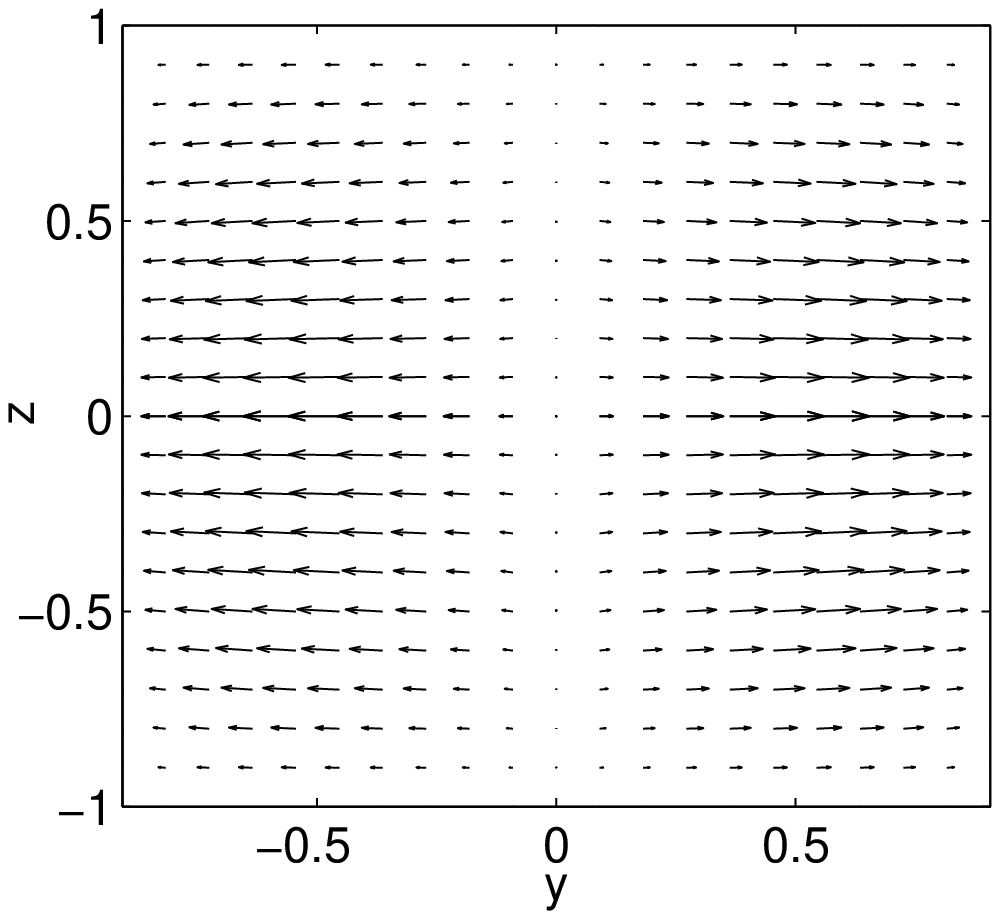}\,\,\,\,\,
\includegraphics[width=.26\textwidth]{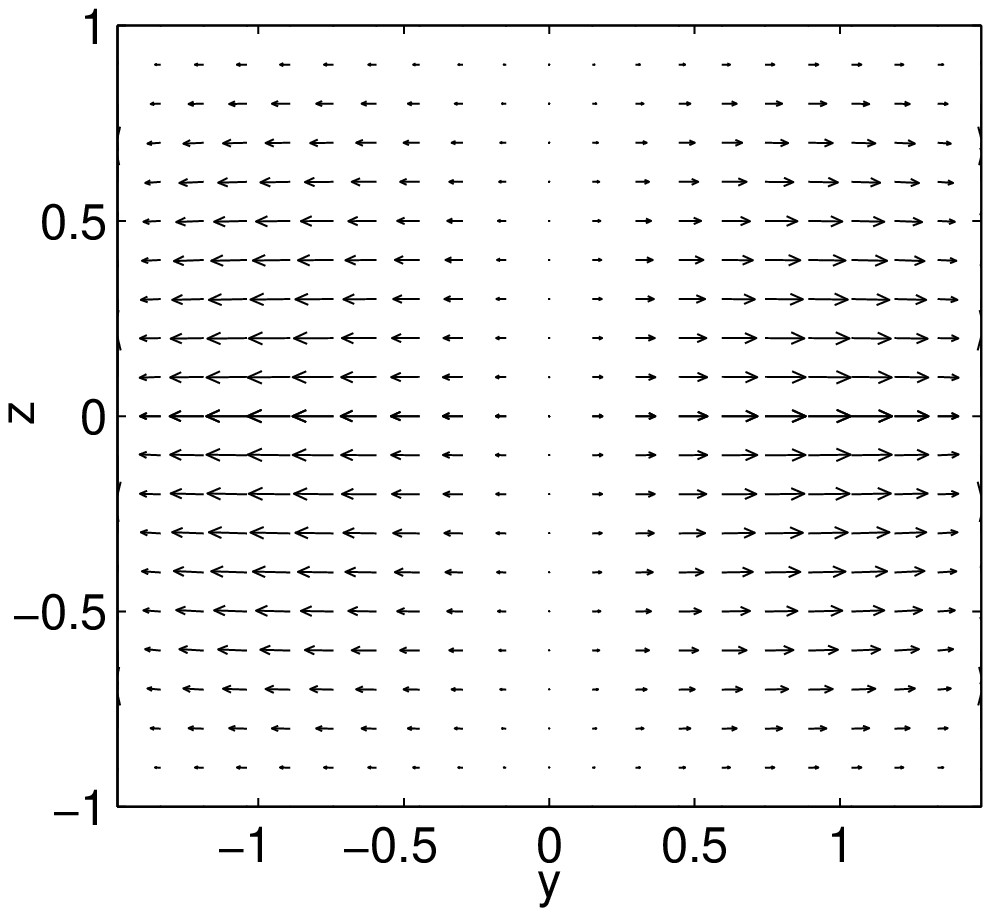}\,\,\,\,\,\,\,\,}
\centerline{
\includegraphics[height=.2\textwidth]{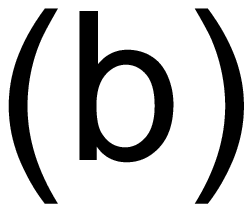}
\includegraphics[width=.26\textwidth]{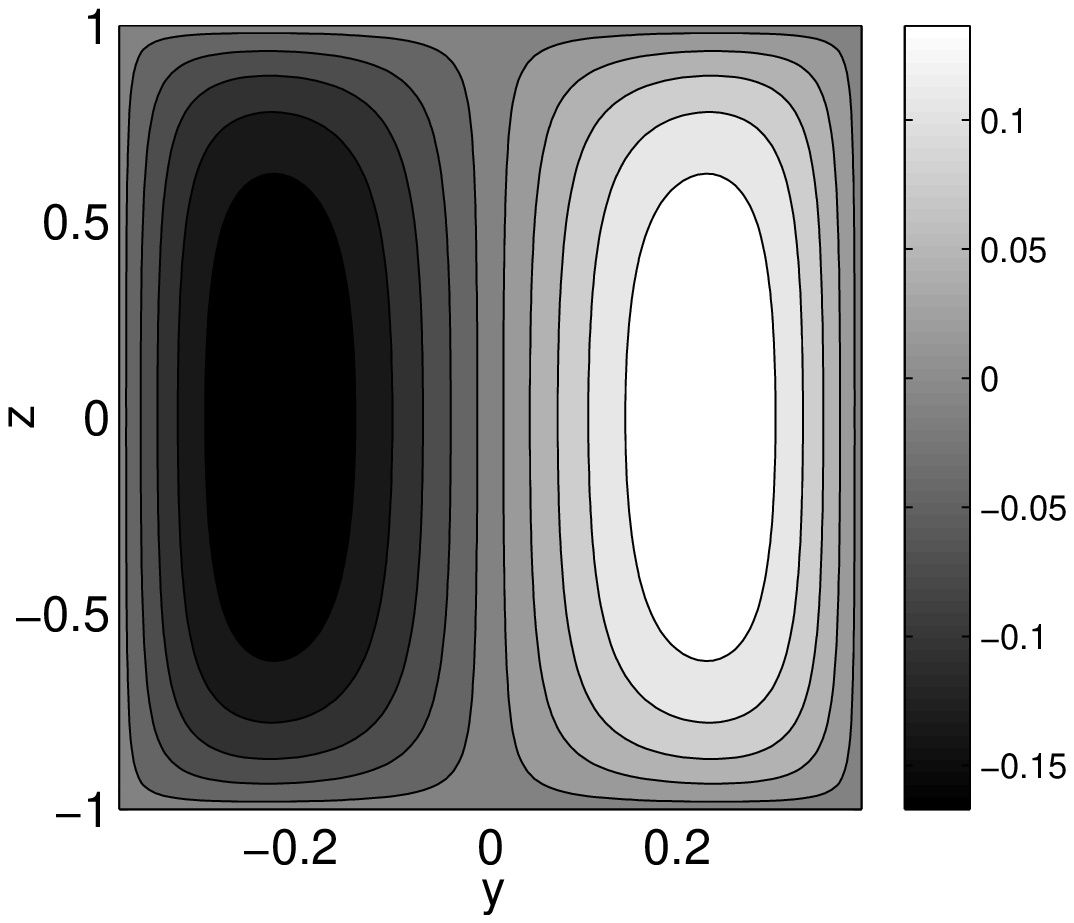}\,\,\,\,
\includegraphics[width=.26\textwidth]{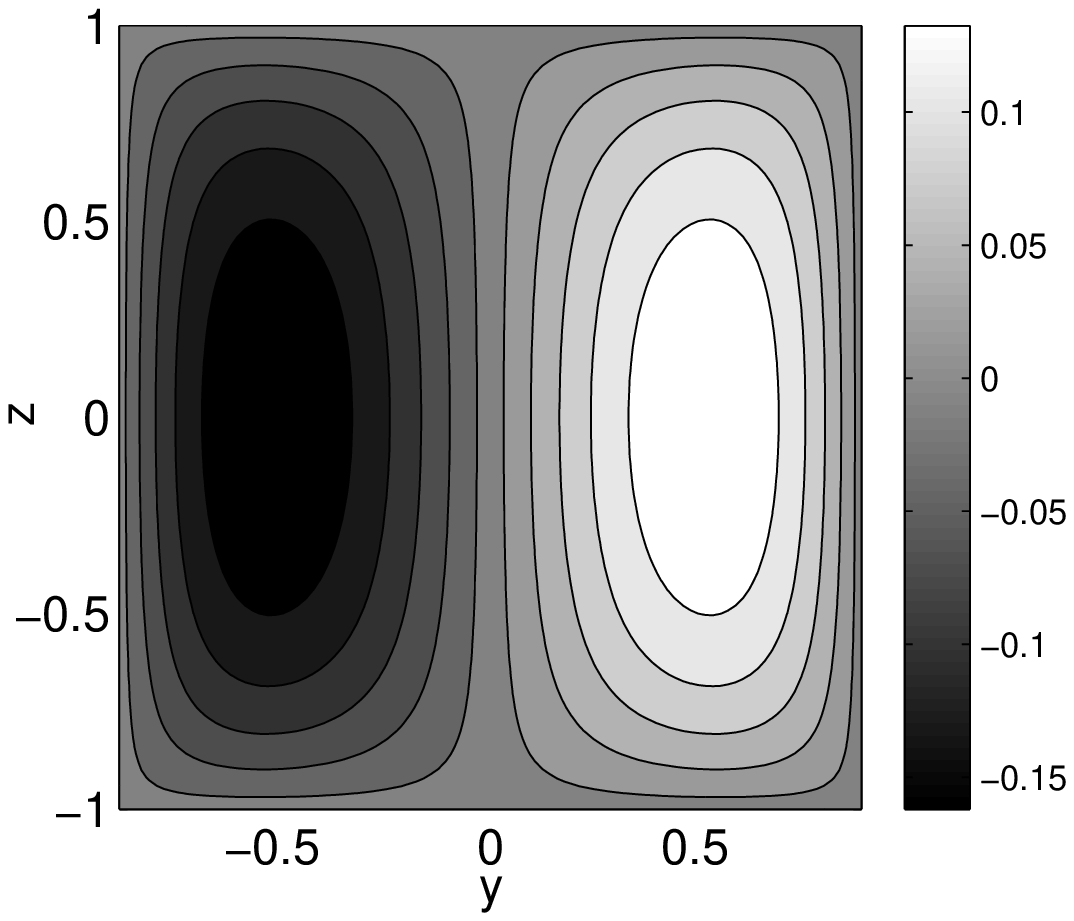}\,\,\,\,
\includegraphics[width=.26\textwidth]{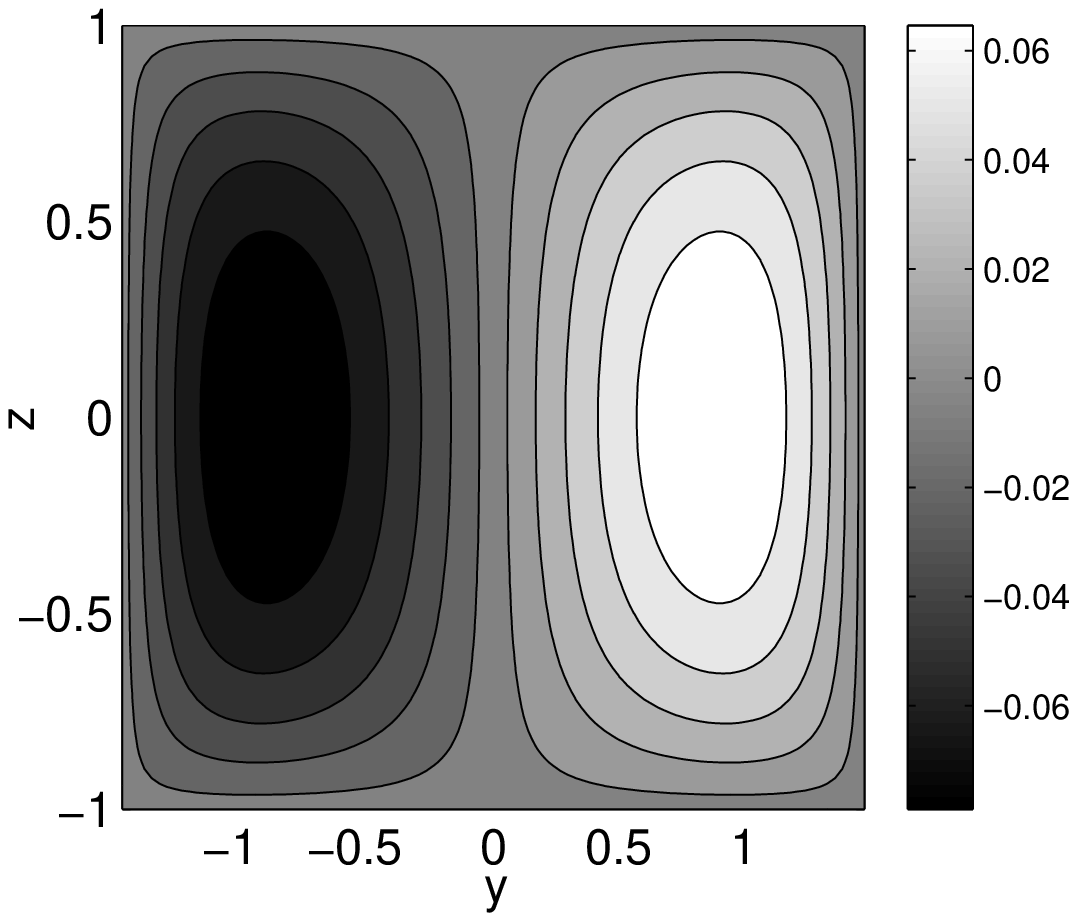}}
\centerline{
\includegraphics[height=.2\textwidth]{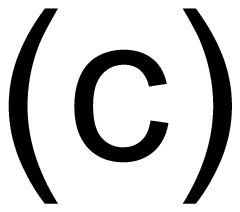}
\includegraphics[width=.26\textwidth]{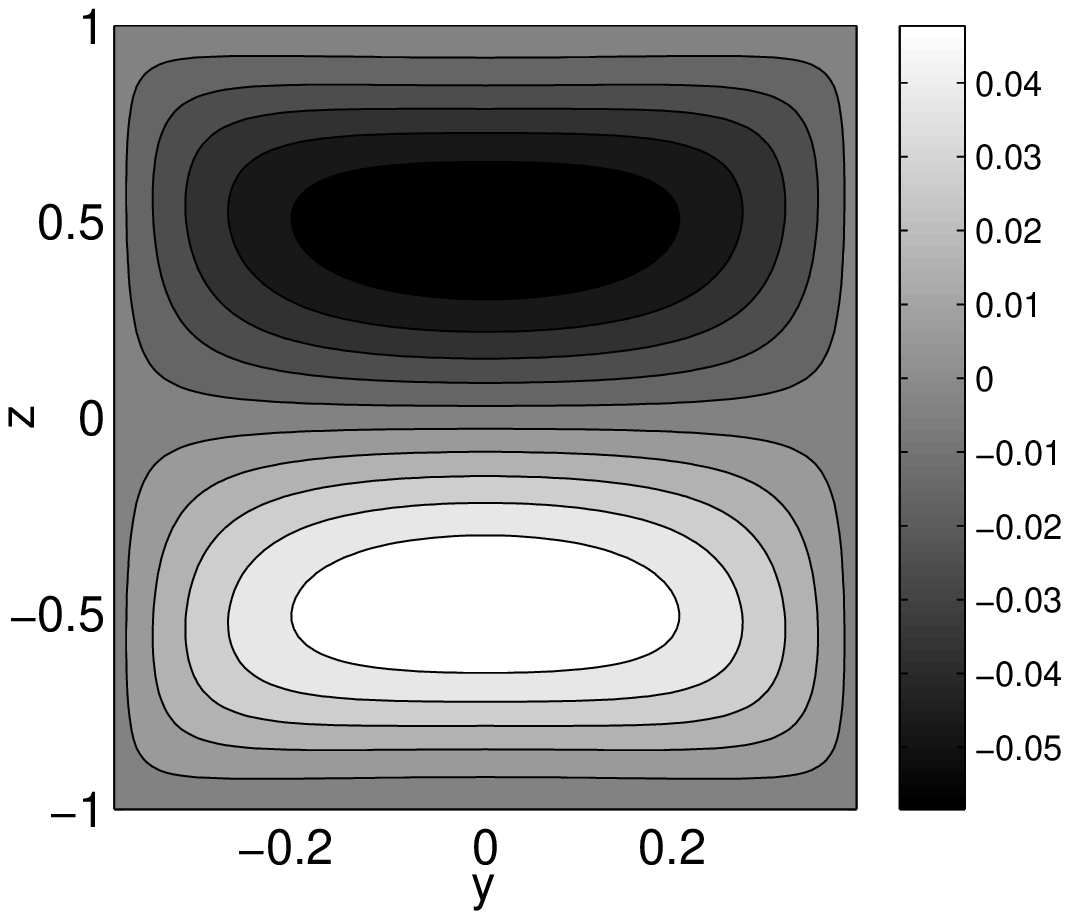}\,\,\,\,
\includegraphics[width=.26\textwidth]{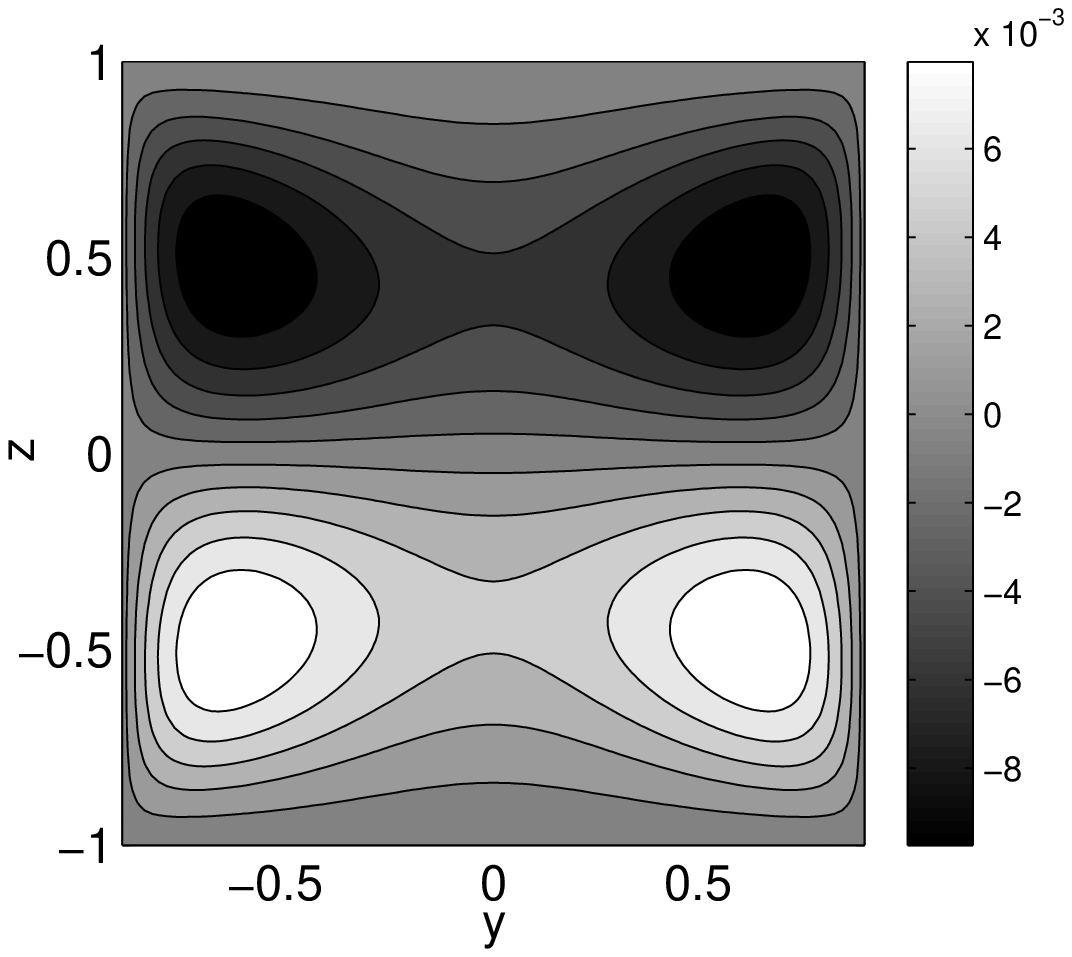}\,\,\,\,
\includegraphics[width=.26\textwidth]{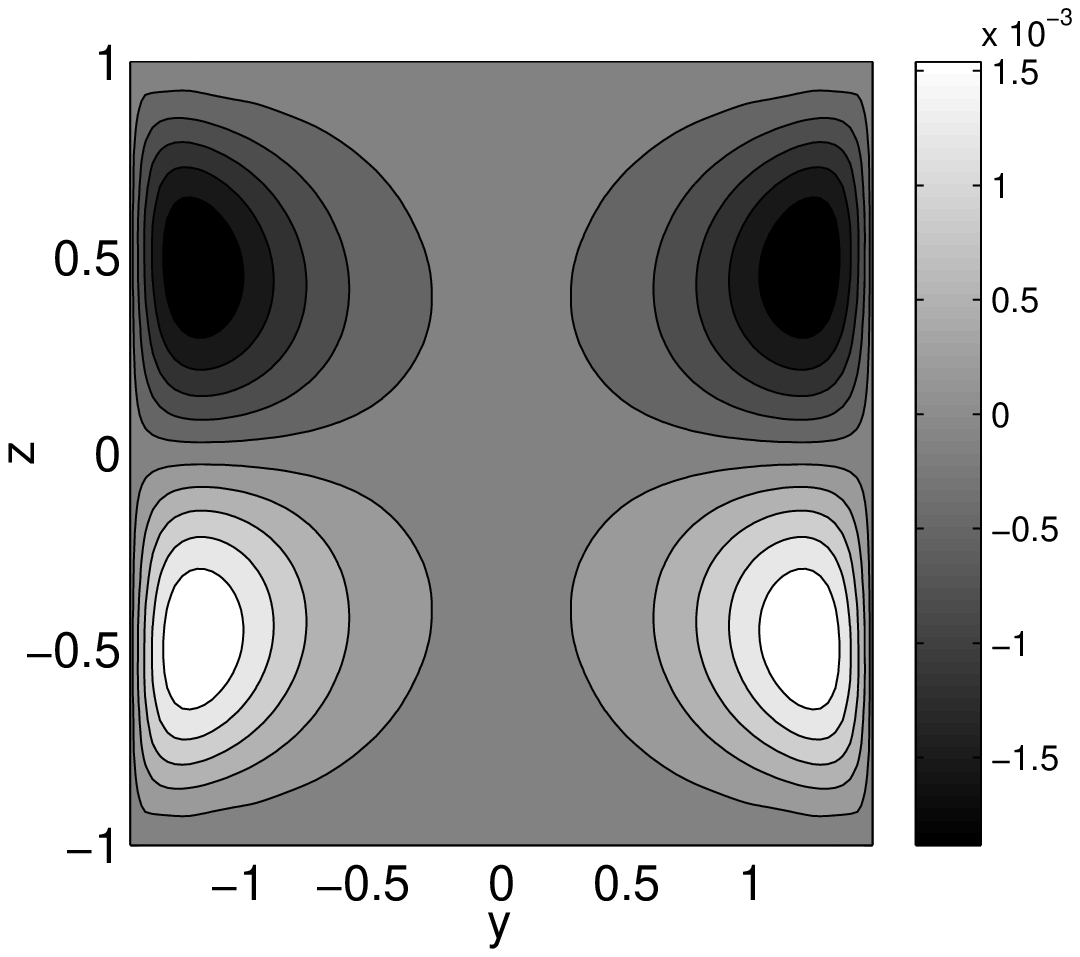}}
\centerline{
\includegraphics[height=.2\textwidth]{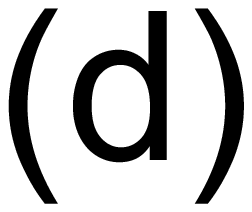}
\includegraphics[width=.26\textwidth]{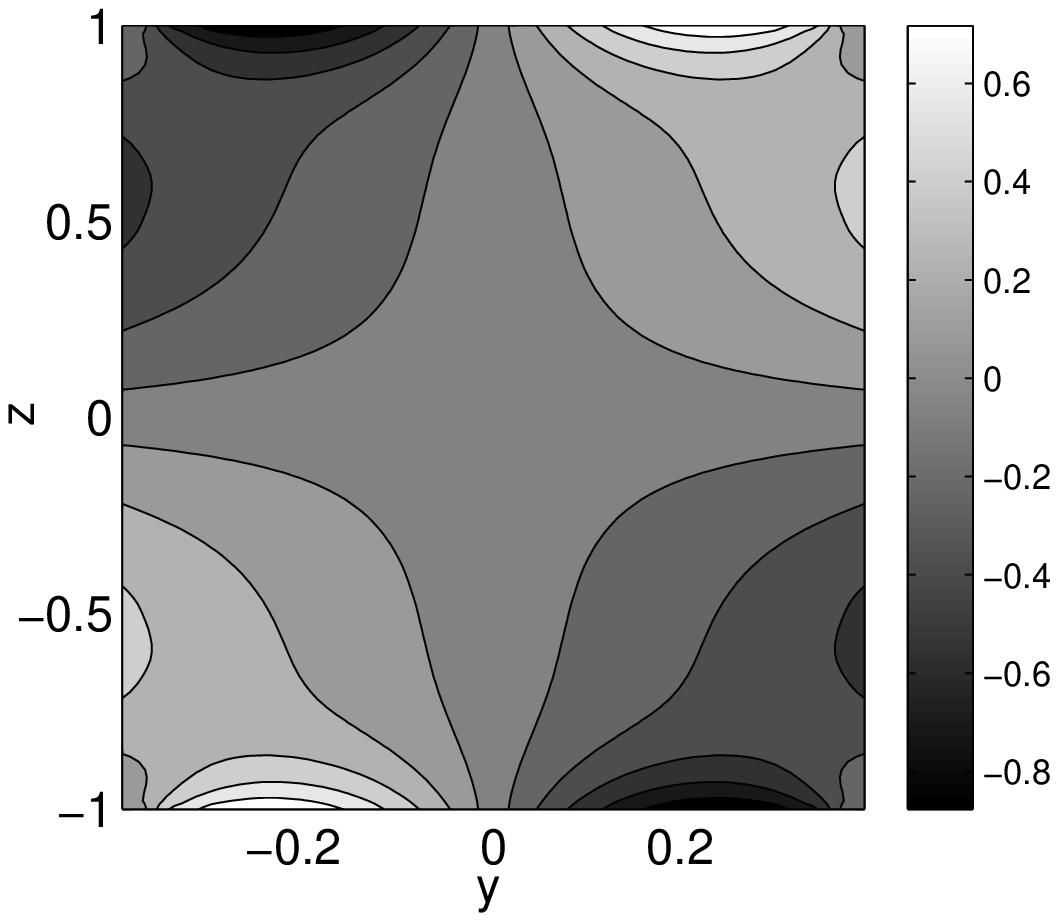}\,\,\,\,
\includegraphics[width=.26\textwidth]{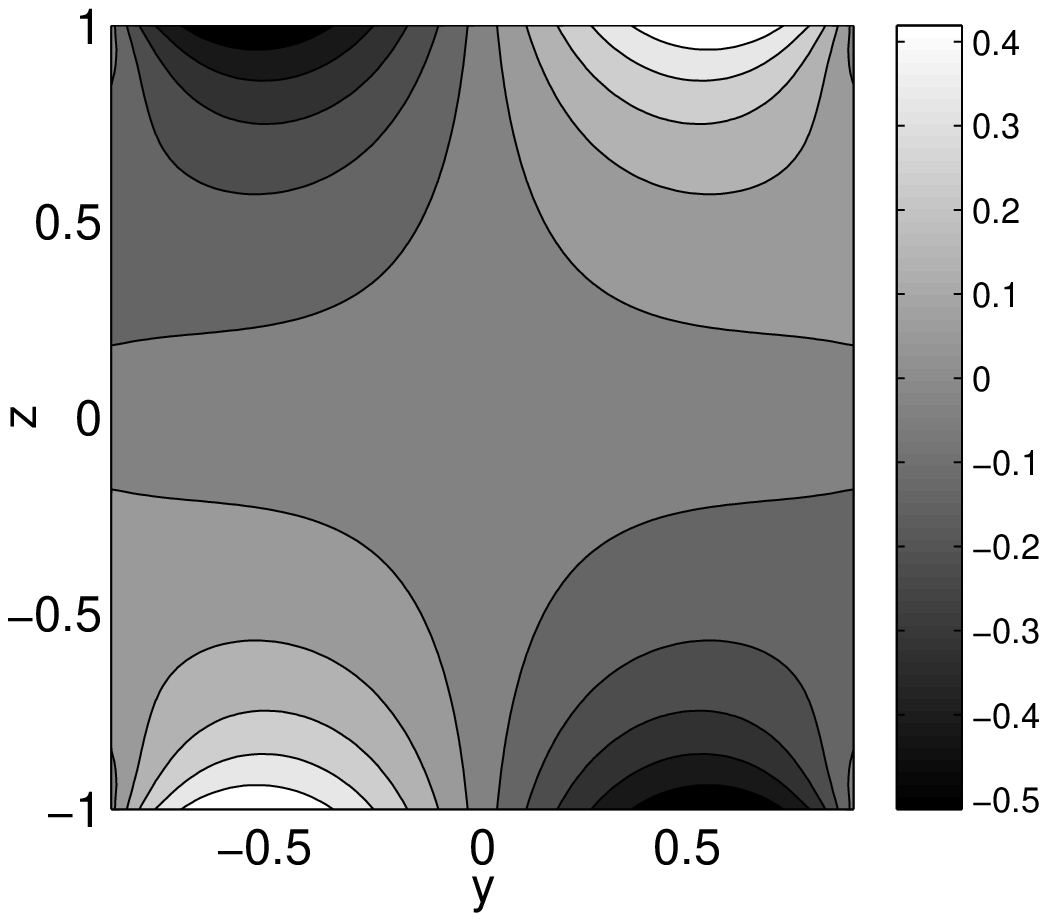}\,\,\,\,
\includegraphics[width=.26\textwidth]{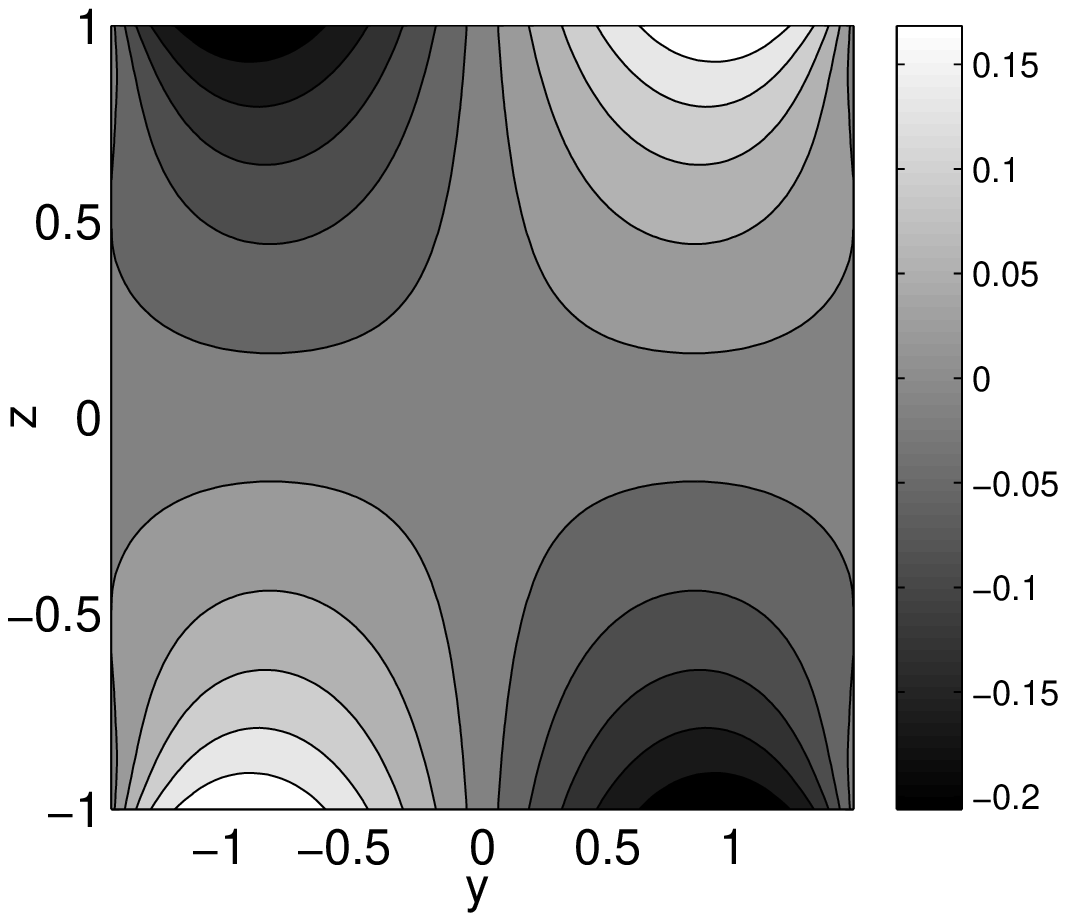}}
\caption{Illustration of the leading-order three-dimensional flow
in the straight planar channel of varying dimensionless
cross-section given by $f(x)=1+0.7\sin x$. Top: view the channel.
Bottom: plots of the leading-order dimensionless cross-sectional
velocity field $(v_0,w_0)$ and axial vorticity $\omega_0$ at three
locations along the channel: $x=5.25$, $6.15$ and $7.05$; (a):
in-plane velocity plots (the velocities are normalized by their
maximum in-plane values); (b): iso-values of the $y$-component
$v_0$ of the velocity, from equation \eqref{v03D}; (c): iso-values
of the $z$-component of the velocity $w_0$, from equation
\eqref{w03D}; (d): iso-values of the $x$-component of the
vorticity $\omega_0$, from equation \eqref{omega03D}.}
\label{quiver}
\end{figure}

\begin{figure}
\centerline{\includegraphics[width=.9\textwidth]{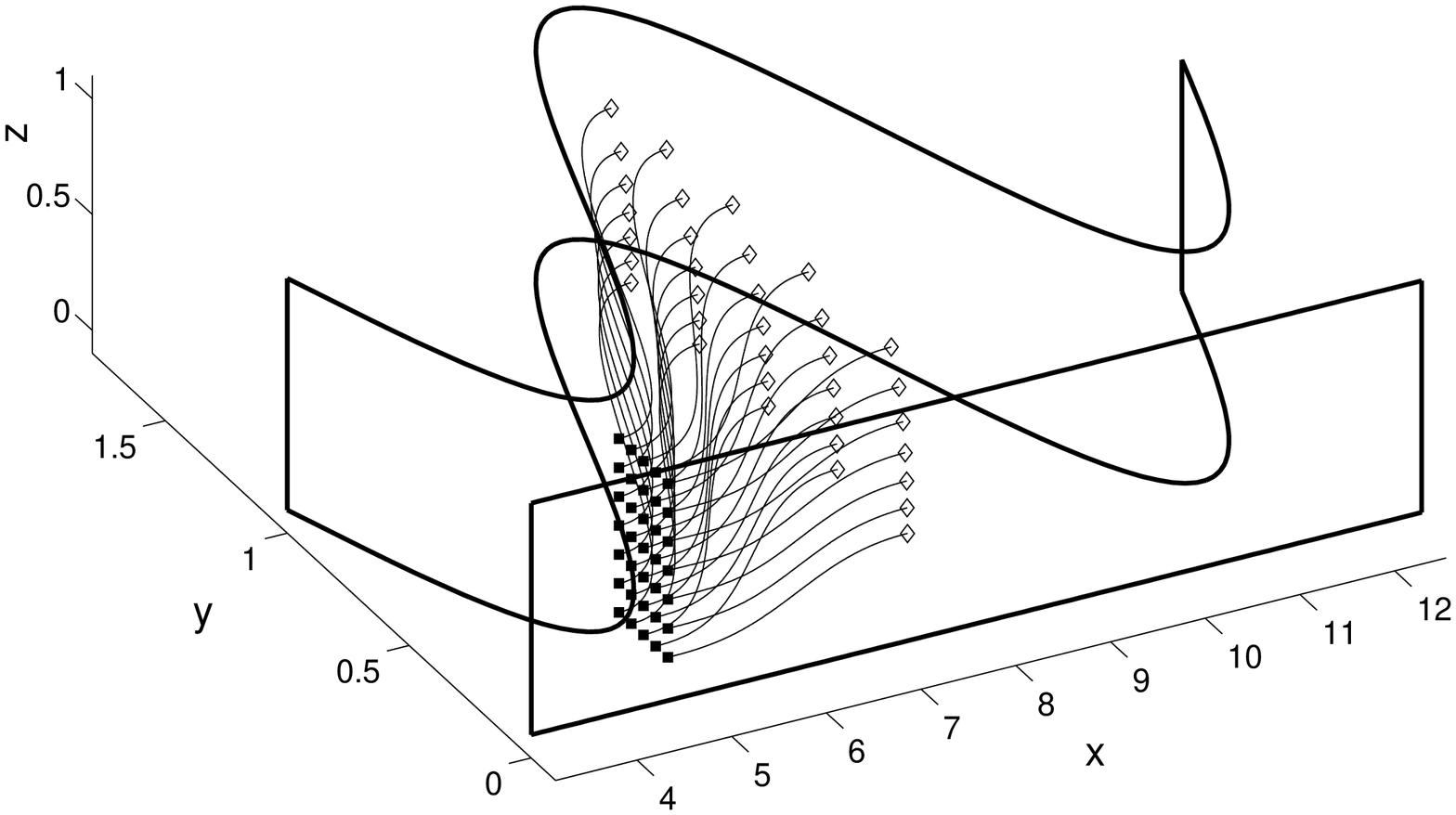}}
\centerline{\includegraphics[width=.5\textwidth]{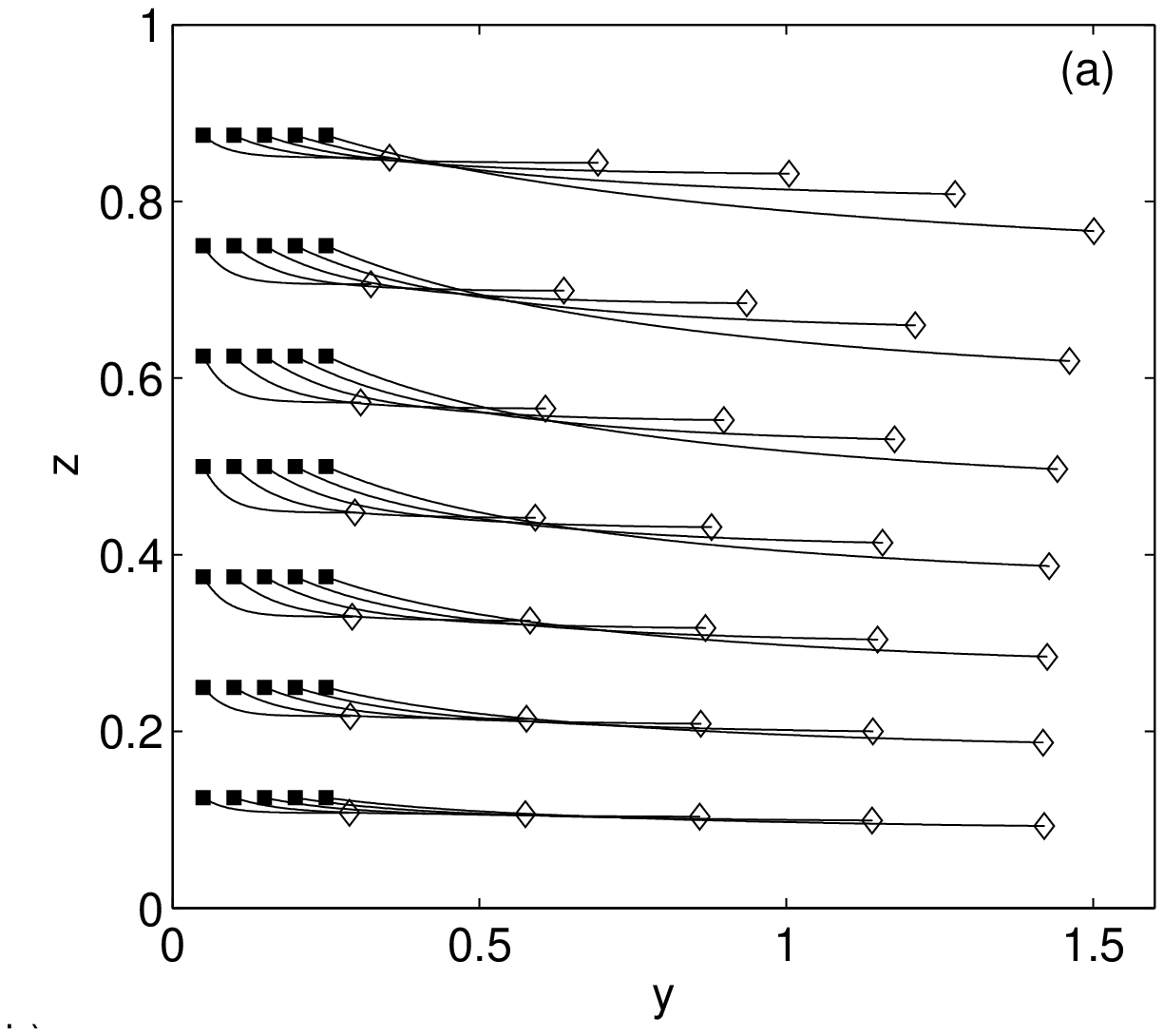}\,\,\,\,\,
\includegraphics[width=.485\textwidth]{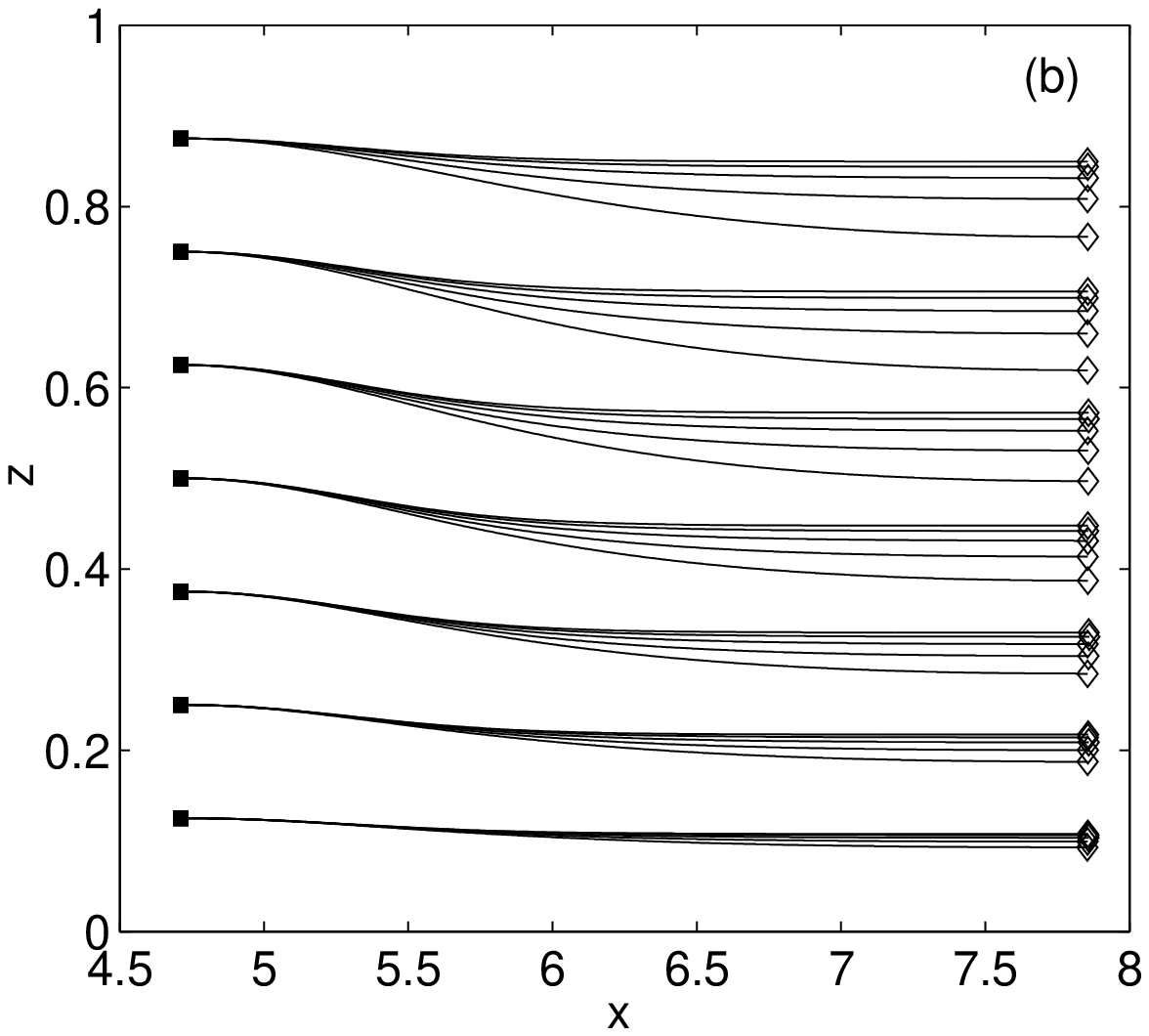}}
\caption{Three-dimensional leading-order streamlines in the planar
channel of varying dimensionless cross-section given by
$f(x)=1+0.7\sin x$. The channel is the same as the one illustrated
in Figure \ref{quiver} and only the streamlines in the quadrant
($y>0,\,z>0$) are reported; those in the other quadrants can be
found using the flow symmetries \eqref{sym}. The dimensionless
time step used for computation is 0.025 and 35 initially evenly
spaced streamlines are considered.  Top: three-dimensional view of
the streamlines between the location of minimum width $x=3\pi/2$
(squares, filled) and the location of maximum width $x=5\pi/2$
(diamonds); the channel boundary and centerplane are also
displayed. Bottom: (a) projection of the streamlines onto the
$(y,z)$ plane; (b) projection of the streamlines onto the $(x,z)$
plane.} \label{particle}
\end{figure}

\begin{figure}
\includegraphics[width=.46\textwidth]{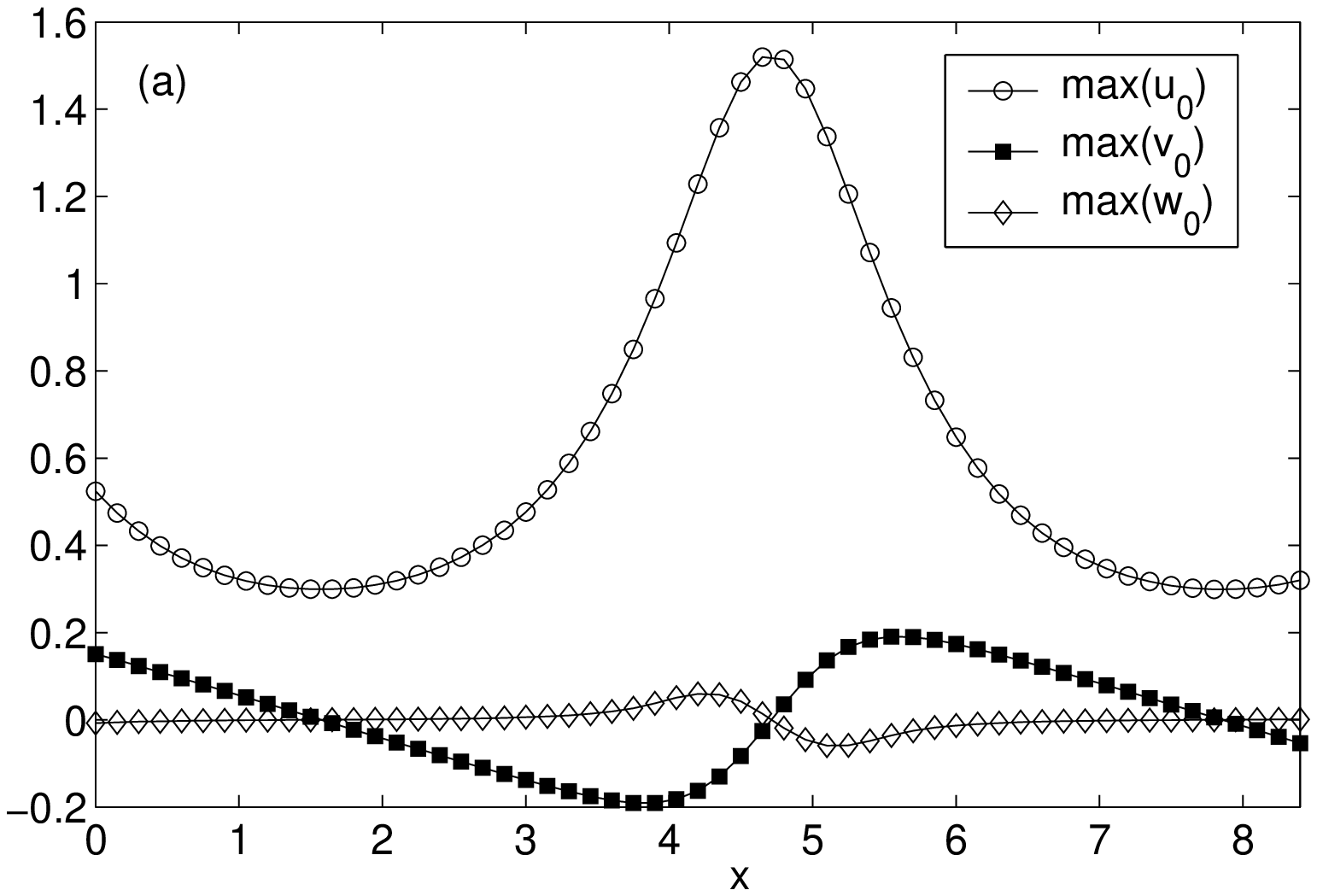}\,\,\,\,
\includegraphics[width=.458\textwidth]{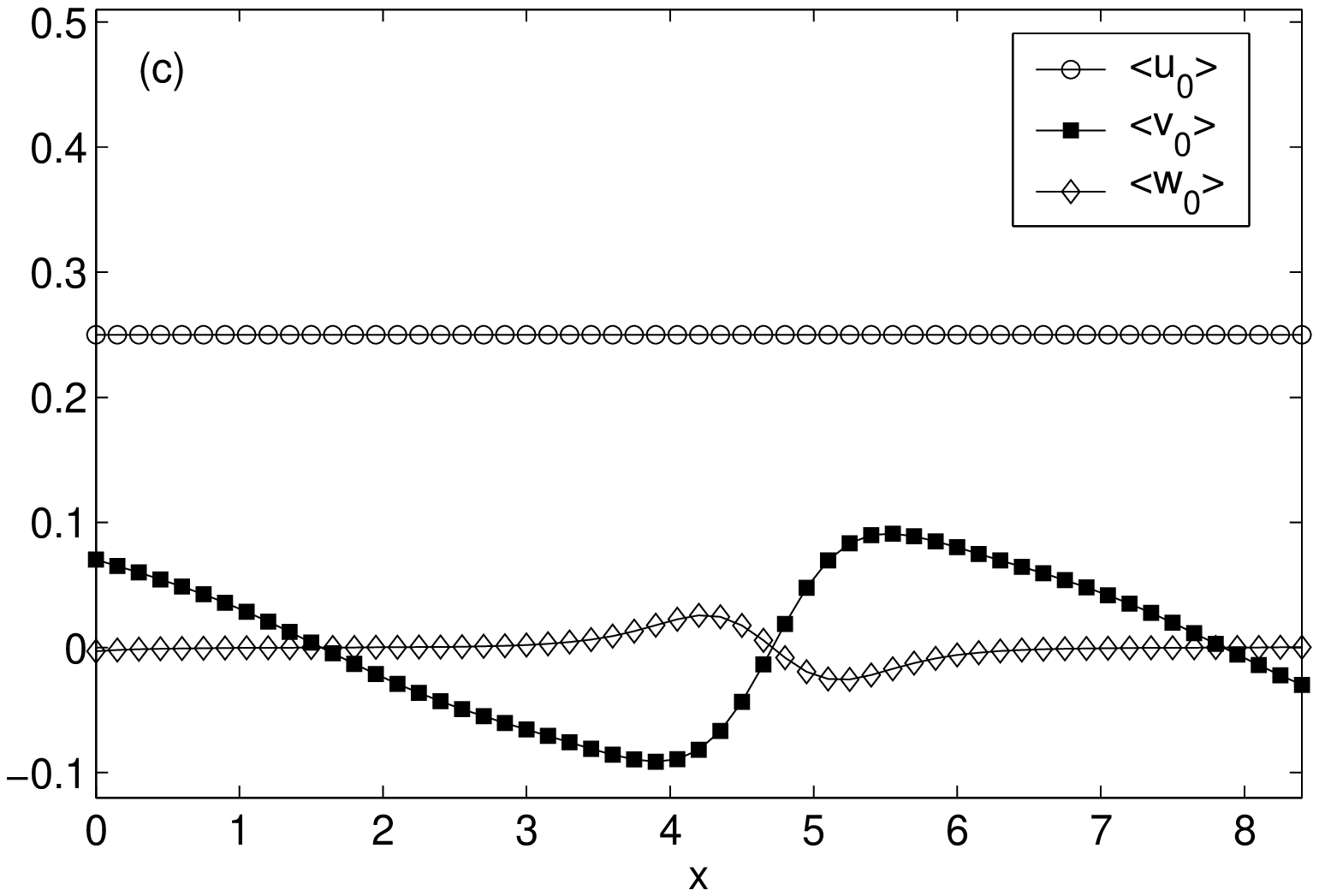}
\includegraphics[width=.46\textwidth]{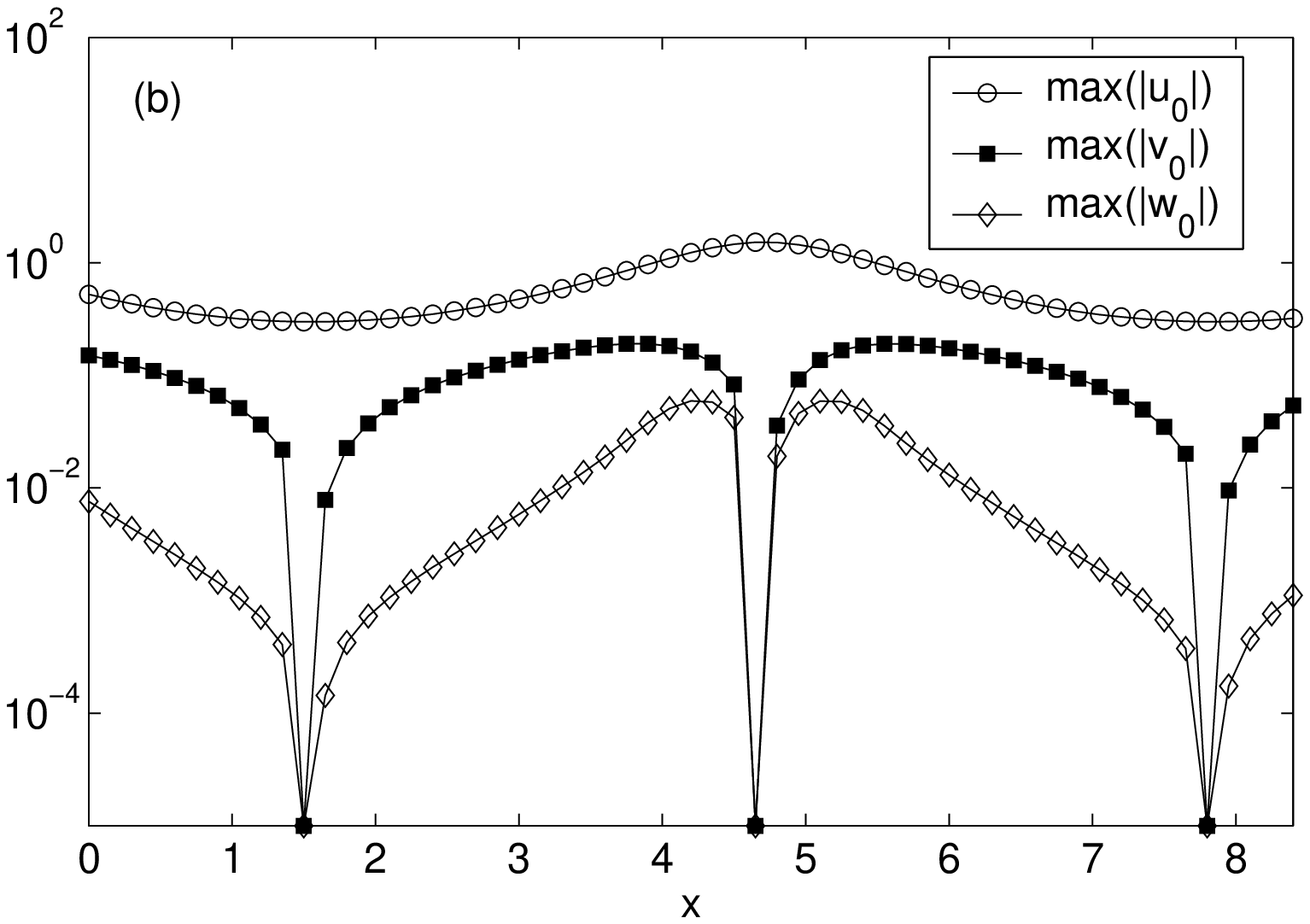}\,\,\,\,\,\,\,
\includegraphics[width=.46\textwidth]{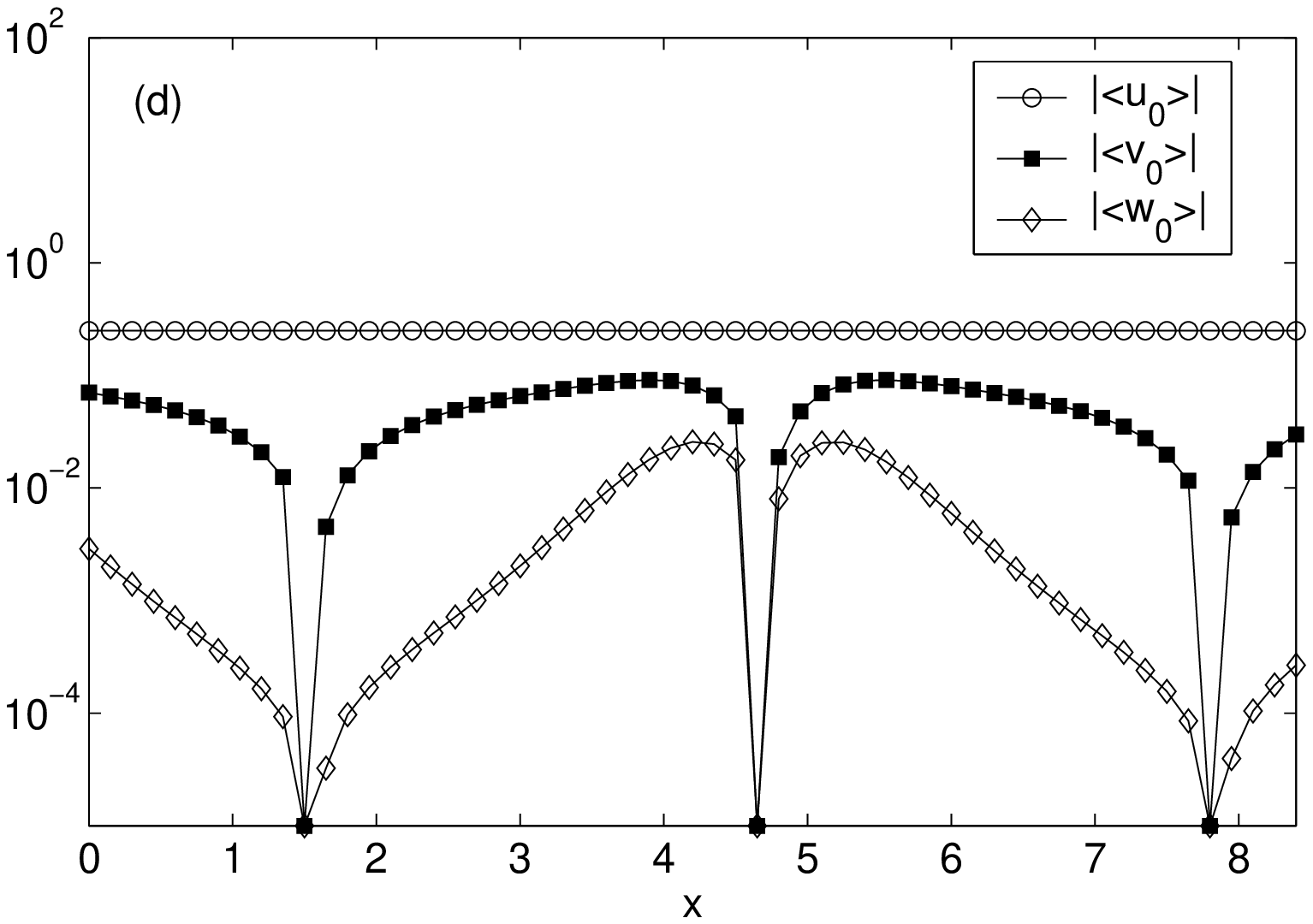}
\caption{Illustration of the leading-order three-dimensional flow
strength in the planar channel of varying dimensionless
cross-section  $f(x)=1+0.7\sin x$. The channel is the same as the
one illustrated in Figure \ref{quiver} and only the velocities in
the quadrant ($y>0,\,z>0$) are considered. Figures (a) and (b):
maximum cross-sectional values of the three components of the
leading-order dimensionless velocity along the channel $u_0$
(circles), $v_0$ (squares, filled) and $w_0$ (triangles); (a):
regular scale, (b): semi-log scale; note that when $v_0$ and $w_0$
were found to be zero, which happens at each location along the
channel where $f'(x)=0$ under the lubrication approximation, they
were replaced by $10^{-5}$ for the semi-log figures. Figures (c)
and (d): same as in (a) and (b) for the quadrant-averaged
velocities $<u_0>$, $<v_0>$ and $<w_0>$; (c): regular scale, (d):
semi-log scale.} \label{max}
\end{figure}

The main results of our flow illustration are displayed in Figures
\ref{quiver}, \ref{particle} and \ref{max}. Figure \ref{quiver}
presents in-plane velocity plots at three locations along the
channel direction, as well as iso-value maps at these locations
for both in-plane velocity components ($v_0$,$w_0$) and for the
axial component of the vorticity ($\omega_0$) \footnote{Note that
the velocity plots in Figure \ref{quiver}a display the
two-dimensional $(y,z)$ secondary flow of a three-dimensional
flow, which is not to be confounded with a two-dimensional flow.}.
Figure \ref{particle} displays the flow streamlines along the
expansion part of the channel ($3\pi/2 < x < 5\pi / 2$). Finally,
Figure \ref{max} displays the maximum cross-sectional as well as
average value of the three components $(u_0,v_0,w_0)$ of the
leading-order dimensionless velocity as a function of the location
along the channel centerline.

The numerical results confirm that the flow at leading-order is
fully three-dimensional. The plots in Figure \ref{quiver} allow us
to visualize the regions of high and low velocity and vorticity
and the streamlines in Figure \ref{particle} show the fluid
elements are indeed vertically displaced as they are advected
along the channel. Note that the similar plots for the contracting
part of the channel were not included here as they can be deduced
from those in Figure \ref{quiver} and \ref{particle} by symmetry
of Stokes's equation.

We also note in Figure \ref{quiver} that the qualitative picture
for the iso-values of $v_0$ do not vary much between the point of
minimum width ($x={3\pi}/{2}$) and the point of maximum width
($x={5\pi}/{2}$). In contrast to $v_0$, the distribution of
vertical velocity $w_0$ is modified appreciably: it changes from a
monotonic variation across the channel (left picture in Figure
\ref{quiver}c) to a variation with local minimum/maximum in the
middle of the channel and global maximum/minimum near the channel
walls (middle and right picture in Figure \ref{quiver}c).
Moreover, as can be seen in Figure \ref{quiver}d, the axial
vorticity is maximum at the top and bottom walls and decays
towards the middle of the channel (z=0); the contracting part is
therefore the position along the channel where the strongest
stirring of material surfaces would occur.

Further, the results of Figure~\ref{max} show that under the
lubrication approximation, the magnitude of the vertical flow
component $w_0$ decreases monotonically during an expansion
$\left({3\pi}/{2}<x<{5\pi}/{2}\right)$; by symmetry of Stokes's
equation, $w_0$ increases in a similar fashion during a
contraction of the channel, $\left({\pi}/{2}<x<{3\pi}/{2}\right)$.

We see also that for the particular case considered here, and
within the lubrication approximation, the leading-order
$y$-component of the velocity $v_0$ is always about one order of
magnitude smaller that the axial component $u_0$ and that the
vertical component $w_0$ is about one order of magnitude smaller
than $v_0$; back to the dimensional variables, these statements
become $v \approx \epsilon u / 10$  and $w \approx \epsilon u /
100$.


Finally, the integrated effect of the vertical flow along the
channel length is illustrated in Figure \ref{particle} by the
vertical deflection of streamlines. The deflection is larger far
from the horizontal centerplane (see Figure \ref{particle}b) and
far from the vertical centerplane (see Figure \ref{particle}a).
Over the channel half period, a fluid element on the upper right
streamline in Figure \ref{particle}a experiences a vertical
displacement of about 10\% of the channel half-height.

\section{Conclusion}
\label{discussion}

We have shown in this paper that the only planar channel shapes
for which the velocity field is two-dimensional under Stokes flow
conditions have both constant curvature and constant cross section
(in which case the flow field is in fact unidirectional). In all
other cases for the variation of the cross-section and curvature,
the velocity is fully three-dimensional at zero Reynolds number
and could in principle be used to mix species in simple
microdevices that can be manufactured with one step of
microfabrication.

A qualitative summary for the occurrence of the third component of
the flow can be given using the two-dimensionality condition, {\it
i.e.} equation \eqref{cond} or \eqref{noslip_curved}. The velocity
field remains two-dimensional in the channel if the
two-dimensional flow rate $Q(z)=\int u \d y $ is constant along
the channel for each $z$. When this is not the case and $Q(z)$ is
streamwise-dependent, a vertical velocity component is induced by
mass conservation. What our study shows is that, under the
lubrication approximation, the only channel geometries with no
embedded obstacles for which this is not the case are those of
both constant width and constant curvature. Note that
alternatively, the presence of obstacles such as cylinders in an
otherwise straight channel would provide similar geometric
features necessary for the occurrence of a three-dimensional flow
\cite{Thompson}.

As the general form of the continuity equation shows, the
magnitude of the ratio of the out-of-plane velocity component $w$
to the axial component $u$ scales as the ratio of the
cross-sectional length scale $h$ to the length scale $\lambda$
over which the variations of the channel geometry occur,
\begin{equation}\frac{w}{u}\approx \frac{h}{\lambda}=\epsilon .
\end{equation}
The numerical results presented in section \ref{illustration} for
a sinusoidal change in cross-section show that the prefactors for
this scaling is about $0.01$ for the ratio of the {\it
leading-order} velocity fields $w_0$ to $u_0$ and indicate poor
mixing. For the case $h \approx \lambda $, we could expect however
all orders in the perturbation expansion to contribute in a non
trivial way, and we expect therefore that with this simple design
a vertical flow of strength comparable to the axial flow could
exist; if that is not the case and the prefactors for the full
calculation are not of order one, the channel will likely present
poor mixing characteristics. Note that as the Reynolds number in
micromixers is not exactly zero but can be as high as 100, we also
expect in this case the occurrence of non-trivial Dean flow-like
contributions to the vertical flow.

We propose to design ''planar mixers'' by a  succession of $n$
mixing cells of length $\lambda$ along a single channel. In each
cell, we expect the integrated displacements $\delta y$ and
$\delta z$ of fluid elements in the cross-section, advected by the
flow at velocity $U_{\rm axial}$, to be given by
\begin{equation}
\delta y \approx \delta z \approx {t_{\perp}}{U_{\perp}},
\end{equation}
where  $t_{\perp}$ is the residence time for the flow in the cell
$t_{\perp}\approx \lambda / U_{\rm axial}$ and $U_{\perp}$ is the
magnitude of the transverse flow, at most $U_{\perp}\approx h
U_{\rm axial} / \lambda$ so that $ \delta y \approx \delta z
\approx h$. Since the total length of the mixer is $n \lambda$ and
the displacements $ \delta y $ and $ \delta z$ are independent of
the cell length, small cells $\lambda \approx h$ should be chosen.
The challenge in the mixing design would then concern (1) the
design of each cell, {\it i.e.} the variations of its radius of
curvature and its cross-section, in order to obtain the maximum
cross-sectional displacement and (2) the setup of the cell
succession in a way that mixing adds up instead of cancelling out;
for example the channel studied in section \ref{illustration}
would obviously make very poor mixing cells because by symmetry of
Stokes flow, every fluid stirring taking place in one part of the
channel would be unstirred in the other part of the channel
located immediately downstream. In general, good performance may
be achieved by avoiding any geometrical symmetry along the
streamwise direction.

The calculations presented in this paper assumed slowly varying
cross-sectional and curvature change along the channel, $\epsilon
\ll 1$. As was shown by Lucas (1972) for two-dimensional channels
of varying shape, the regular perturbation expansions \eqref{exp}
or \eqref{exp2} are expected to have have order one or larger
radius of convergence in $\epsilon$; as a consequence, the
conclusions reached using the leading-order velocity fields are
valid for the entire velocity field when $\epsilon$ is ${\cal
O}(1)$, and presumably higher even though our results cannot be
applied directly. With current microfabrication techniques, the
minimum in-plane dimension ($\approx\lambda$) that can be
generated is typically greater than the minimum out-of-plane
dimension ($\approx h$). As a consequence, the cross-sectional
dimensions of microchannels tend to satisfy the criterion,
$\epsilon < 1$, and the results obtained in this paper are
expected to apply for all such cases.

The limitation of the passive mixing strategy proposed here lies
in the top-bottom symmetry for the velocity field,
$(u,v,w)(x,y,-z)=(u,v,-w)(x,y,z)$, due to the symmetries of the
Stokes equations. Mixing can therefore not be achieved between the
fluids located in the $z>0$ and $z<0$ planes and consequently, the
streams of solutions that are to be mixed must be introduced at
the inlet of the channel with alignment in the vertical direction.
The case of a straight channel of varying section studied in
section \ref{straight} also possess a right/left symmetry,
$(u,v,w)(x,-y,z)=(u,-v,w)(x,y,z)$, (see Figure \ref{quiver}) and
therefore cannot mix species fluids located in the $y>0$ and $y<0$
planes. The configuration studied in section \ref{curved} does not
possess such a symmetry and should be used to transport fluid
between the $n>0$ and $n<0$ planes, similarly to what was achieved
in Stroock {\it et al.} (2002). A fully numerical approach to the
problem (using e.g. a boundary element method or a commercial
code) would allow a detailed study of the proposed mixing design,
its optimization, and dispersion characteristics.

\begin{acknowledgments}
The authors acknowledge fruitful discussions with M. Brenner and
T. Squires on the subject. We also thank two anonymous referees
for useful suggestions. Funding from the Harvard MRSEC and the NSF
Division of Mathematical Sciences is gratefully acknowledged.
\end{acknowledgments}

\appendix
\section{Proofs}
\subsection{Proof of result \eqref{constant_straight} for straight channels}
\label{A1}

We show in this section that the only set of width functions
$f(x)$ that satisfy \eqref{constant_straight} are the constant
functions. Let us assume that \eqref{constant_straight} is
satisfied for a function $f(x)$. We first rewrite equation
\eqref{gradp} for the pressure gradient under the form
\begin{equation}
6\frac{\d p_0}{\d x} \sum_{n\geq 0}\frac{\tanh (k_nf(x))}{k_n^5}
-f(x)\frac{\d p_0}{\d x} =\frac{3}{4}\cdot \label{ps1}
\end{equation}
We then rewrite the condition \eqref{constant_straight} for
two-dimensionality of the flow as \be \frac{\d p_0}{\d x} \tanh
(k_nf(x))=\delta_n+k_n \frac{\d p_0}{\d x} \label{ps2}
\end{equation}
Substituting the expression obtained in \eqref{ps2} into
\eqref{ps1} leads to a closed form solution for the axial pressure
gradient
\begin{equation}
\frac{\d p_0}{\d x} =\frac{\Delta_1}{\Delta_2-f(x)} ,\quad
\Delta_1=\frac{3}{4}-6\sum_{n\geq 0}\frac{\delta_n}{k_n^5} ,\quad
\Delta_2=6\sum_{n\geq 0}\frac{1}{k_n^4}\cdot \label{func1}
\end{equation}
As a conclusion, the functional form \eqref{func1} obtained for
the pressure gradient is not consistent with that given by the
assumption of the two-dimensionality of the flow
\eqref{constant_straight} \be \frac{\d p_0}{\d x}
=\frac{\delta_n}{\tanh (k_n f(x)) -k_n},
\end{equation}
unless the function $f(x)$ is constant.

\subsection{Proof of result \eqref{constant_curved} for curved channels}
\label{A2} We show in this section that the only set of curvature
functions $f(s)$ that satisfy \eqref{constant_curved} are the
constant functions. Let us assume that \eqref{constant_curved} is
satisfied for a function $f(s)$. We first rewrite equation
\eqref{pressure_curved} for the pressure gradient under the form
\begin{equation}
2f(s)\frac{\d p_0}{\d s} \sum_{n\geq 0}(-1)^n\frac{\alpha
H_n(s)}{k_n^2} -\frac{2f(s)}{3\beta}\frac{\d p_0}{\d s} \ln
\left(\frac{f(s)+\beta}{f(s)-\beta}\right) =1. \label{p1}
\end{equation}
We then rewrite the condition \eqref{constant_curved} for
two-dimensionality of the flow as
\begin{equation}
f(s)\frac{\d p_0}{\d s} \frac{\alpha
H_n(s)}{k_n}=\frac{2(-1)^n}{\beta
k_n^3\ln\left(\frac{f(s)+\beta}{f(s)-\beta}\right)}f(s)\frac{\d
p_0}{\d s} +\gamma_n. \label{p2}
\end{equation}
Substituting \eqref{p2} into \eqref{p1} leads to a closed-form
solution for the streamwise pressure gradient
\begin{equation}
f(s)\frac{\d p_0}{\d
s}=\frac{\Delta_3\ln\left(\frac{f(s)+\beta}{f(s)-\beta}\right)}
{\ln^2\left(\frac{f(s)+\beta}{f(s)-\beta}\right)-\Delta_4} ,\quad
\Delta_3=3\beta\left\{\sum_{n\geq
0}\frac{(-1)^n\gamma_n}{k_n}-\frac{1}{2}\right\} ,\quad
\Delta_4=6\sum_{n\geq 0}\frac{1}{k_n^4}\cdot \label{func2}
\end{equation}
Further, it is possible to use the asymptotic behaviors near
$x\sim \infty$ of Bessel functions (see e.g. \cite{Abra}) $ I_p
(x)\sim e^x / \sqrt{2\pi x}$, $K_q(x)\sim {\pi e^{-x}}/{\sqrt{2
x}}$ to obtain the asymptotic behaviors of $E_n(s)$, $F_n(s)$,
$G_n(s)$ as $n\rightarrow+\infty$, from equations \eqref{En},
\eqref{Fn} and \eqref{Gn} respectively. It is then straightfoward
to obtain the asymptotic behavior of $H_n$
\begin{equation}
H_n(s)\sim \frac{4(-1)^nf(s)}{k_n^3(f(s)^2-\beta^2)}\cdot
\end{equation}
This behavior and the condition of two-dimensionality
\eqref{constant_curved} allows to obtain an alternate functional
behavior for the streamwise pressure gradient
\begin{equation}
 f(s)\frac{\d
p_0}{\d s}\sim  \Delta_5
\ln\left(\frac{f(s)+\beta}{f(s)-\beta}\right),\quad
\Delta_5=\frac{\beta k_n^3 (-1)^{n+1}\gamma_n}{2}\cdot
\label{func3}
\end{equation}
As a conclusion, the two functional forms obtained assuming
two-dimensionality of the flow \eqref{func2} and \eqref{func3} are
not consistent with each other unless the function $f(s)$ is a
constant.

\section{Leading order velocity field for asymmetric wall profile}
\label{asym}
\begin{figure}
\centering
\includegraphics[width=.7\textwidth]{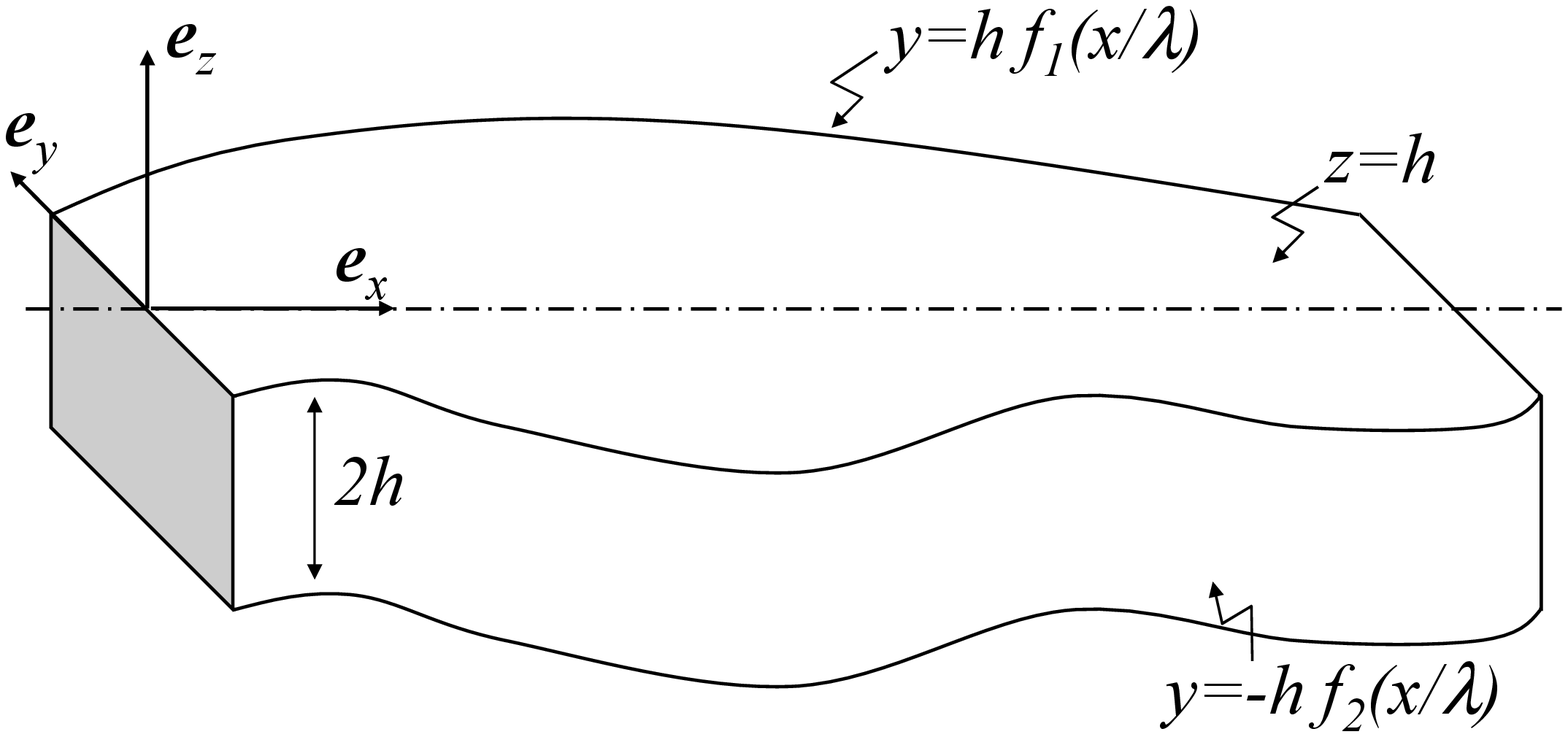}
\caption{General shape of a (asymmetric) planar channel.}
\label{fig3}
\end{figure}

We present in this section the solution for the leading-order
lubrication velocity field in the case of a planar channel of
general shape; the calculation presented here contains that
presented in section \ref{solution} as a special case. We use a
cartesian coordinate system, as illustrated in Figure \ref{fig3}.
All notations  are the same as in section \ref{straight}, with the
difference that the channel is not right/left symmetric and we
denote its boundaries by the equations $y=hf_1(x/\lambda)$ and
$y=-hf_2(x/\lambda)$, with $f_1(x)>0$ and $f_2(x)>0$. Using the
same nondimensionalizations and regular expansion as in section
\ref{straight} and defining the convenient  notations
$f_1\triangleq f_1(x)$ and $f_2\triangleq f_2(x)$, we obtain that
the leading-order axial velocity field is given by
\begin{equation}
u_0(x,y,z)=\sum_{n\geq 0}{\cal U}_n(x,y)\cos k_n z, \quad {\cal
U}_n(x,y)=2\frac{\d p_0}{\d
x}\frac{(-1)^n}{k_n^3}\left\{\frac{\cosh
k_n\left(y+\frac{f_2-f_1}{2}\right)}{\cosh
k_n\left(\frac{f_1+f_2}{2}\right)}-1\right\},
\end{equation}
and a pressure gradient obtained by enforcing the flow rate
condition
\begin{equation}
2 \int_{0}^{1} \int_{-f_2}^{f_1} u_0 \, \d y\, \d z =1,
\end{equation}
which leads to
\begin{equation} \frac{\d p_0}{\d x} =\frac{3}{2(f_1+f_2)}\left\{
\frac{12}{f_1+f_2}\sum_{n\geq 0}\left(\frac{\tanh
\left(k_n\frac{f_1+f_2}{2}\right)}{k_n^5}\right) -1
\right\}^{-1}\cdot
\end{equation}
Concerning the other two velocity components, the arguments
presented in section \ref{subset} also apply to the configuration
considered here and the equations to solve for $v_0$ and $w_0$ are
equations \eqref{biv} and \eqref{cont3D}, along with the no-slip
boundary conditions. Using the top/bottom symmetries in the
velocity field
\begin{equation}
v_0(x,y,-z)  =   v_0(x,y,z),\quad w_0(x,y,-z)  = - w_0(x,y,z)
\end{equation}
we look for for a solution of \eqref{biv} under the form of a
double Fourier series in $y$ and $z$,
\begin{eqnarray}
\label{vgeneral} v_0(x,y,z)&=&\sum_{n\geq 0} A_n(x,y) \cos k_nz
+\sum_{m> 0}
B_m(x,z) \sin\left( \frac{\ell_m(2y+f_2-f_1)}{f_1+f_2}\right)\\
&& +\sum_{p\geq 0} C_p(x,z) \cos\left( \frac{k_p
(2y+f_2-f_1)}{f_1+f_2}\right),\nonumber
\end{eqnarray}
where $k_n=(n+1/2)\pi$, $\ell_m=m\pi$, $k_p=(p+1/2)\pi$ and
\begin{subeqnarray}
A_n(x,y)&=&{a}_n(x)P_n(x,y)+b_n(x)Q_n(x,y), \\
B_m(x,z)&=&c_m(x)S_m(x,z),\\
C_p(x,z)&=&d_p(x)T_p(x,z),
\end{subeqnarray}
with
\begin{subeqnarray}
P_n(x,y)&=&(f_1+f_2)\sinh  k_n(f_1-y)-(f_1-y)\cosh k_n(f_1-y)
\tanh k_n(f_1+f_2),\,\,\,\,\,\,\,\,\, \\
Q_n(x,y)&=&(f_1-y)\sinh k_n(f_1-y)-(f_1-y)\cosh k_n(f_1-y) \tanh
k_n(f_1+f_2),\\
S_m(x,z)&=&\cosh\left(\frac{2\ell_mz}{f_1+f_2}\right)
-z\frac{\sinh\left(\frac{2\ell_mz}{f_1+f_2}\right)}{\tanh\left(\frac{2\ell_m}{f_1+f_2}\right)},\\
T_p(x,z)&=&\cosh\left(\frac{2k_pz}{f_1+f_2}\right)
-z\frac{\sinh\left(\frac{2k_pz}{f_1+f_2}\right)}{\tanh\left(\frac{2k_p}{f_1+f_2}\right)}\cdot
\end{subeqnarray}
Note that the solution in \eqref{vgeneral} satisfies the no-slip
boundary condition on all four walls. The integration of the
continuity equation \eqref{cont3D} leads to the general form of
the vertical velocity component
\begin{eqnarray}
\label{w0general} w_0(x,y,z)&=&\sum_{n\geq
0}\left\{-\frac{1}{k_n}\left(\frac{\p {\cal U}_n}{\p
x}+a_n(x)\frac{\p P_n}{\p y}+b_n(x)\frac{\p Q_n}{\p y}
\right) \right\}\sin k_nz\\
&& +\sum_{m> 0}\ell_m
c_m(x)Y_m(x,z)\cos\left(\frac{\ell_m(2y+f_2-f_1)}{f_1+f_2}\right)\nonumber\\
&& +\sum_{p\geq 0}k_p
d_p(x)Z_p(x,z)\sin\left(\frac{k_p(2y+f_2-f_1)}{f_1+f_2}\right)\cdot
\nonumber\end{eqnarray} with
\begin{subeqnarray}
Y_m(x,z)&=&-\frac{1}{\ell_m}\left\{\left(1+\frac{1}{\left(
\frac{2\ell_m}{f_1+f_2}\right) \tanh \left(
\frac{2\ell_m}{f_1+f_2}\right)} \right)\sinh\left(
\frac{2\ell_mz}{f_1+f_2}\right)-\frac{z\cosh\left(
\frac{2\ell_mz}{f_1+f_2}\right) }{\tanh \left(
\frac{2\ell_m}{f_1+f_2}\right)} \right\},\\
Z_p(x,z)&=&\frac{1}{k_p}\left\{\left(1+\frac{1}{\left(
\frac{2k_p}{f_1+f_2}\right) \tanh \left(
\frac{2k_p}{f_1+f_2}\right)} \right)\sinh\left(
\frac{2k_pz}{f_1+f_2}\right)-\frac{z\cosh\left(
\frac{2k_pz}{f_1+f_2}\right) }{\tanh \left(
\frac{2k_p}{f_1+f_2}\right)} \right\}\cdot
\end{subeqnarray}
The sets of unknown functions $(a_n)$, $(b_n)$, $(c_n)$ and
$(d_n)$ are finally determined by enforcing the no-slip boundary
condition for $w_0$ in \eqref{w0general} on the four walls $z=\pm
1$, $y=f_1$ and $y=-f_2$, and we obtain a set of four infinite
linear algebraic equations
\begin{equation}\left\{
\begin{array}{rcl}
a_i(x) & = & \displaystyle\overline{a_i}(x) +\sum_{n\geq
0}A^{(2)}_{in}(x) b_n(x) +\sum_{m> 0}A^{(3)}_{im}(x) c_m(x)
+\sum_{p\geq 0}A^{(4)}_{ip}(x) d_p(x)\,\,\,\,(i\geq 0),\\
b_n(x) & = & \displaystyle\overline{b_n}(x) +\sum_{i\geq
0}B^{(1)}_{ni}(x) a_i(x) +\sum_{m > 0}B^{(3)}_{nm}(x) c_m(x)
+\sum_{p\geq 0}B^{(4)}_{np}(x) d_p(x)\,\,\,\,(n\geq 0),\\ c_m(x) &
= & \displaystyle\overline{c_m}(x)+\sum_{i\geq 0}C^{(1)}_{mi}(x)
a_i(x)+\sum_{n\geq 0}C^{(2)}_{mn}(x)
b_n(x)+\sum_{p\geq 0}C^{(4)}_{mp}(x) d_p(x)\,\,\,\,(m> 0),\\
d_p(x) & = & \displaystyle\overline{d_p}(x) +\sum_{i\geq
0}D^{(1)}_{pi}(x) a_i(x) +\sum_{n\geq 0}D^{(2)}_{pn}(x)
b_n(x)+\sum_{m> 0}D^{(3)}_{pm}(x) c_m(x) \,\,\,\,(p \geq 0),
\label{dual2}
\end{array}\right.
\end{equation}
with
\begin{subeqnarray}
\displaystyle\overline{a_i}(x) & = & -\left(\frac{\p P_i}{\p y}(x,f_1(x)) \right)^{-1} \frac{\p u_i}{\p x}(x,f_1(x)),\\
\displaystyle\overline{b_n}(x) & = & -\left(\frac{\p Q_n}{\p y}(x,-f_2(x)) \right)^{-1} \frac{\p {\cal U}_n}{\p x}(x,-f_2(x)),\\
\displaystyle\overline{c_m}(x) & = &
\frac{2}{\ell_mY_m(x,1)(f_1+f_2)}\left\{\int_{-f_2}^{f_1}\left(
\sum_{j\geq 0}\frac{(-1)^j}{k_j} \frac{\p u_j}{\p x} (x,y)\right)
\cos\left(\frac{\ell_m(2y+f_2-f_1)}{f_1+f_2}\right)\d y
\right\}, \\
\displaystyle\overline{d_p}(x) & = & \frac{2}{k_p
Z_p(x,1)(f_1+f_2)}\left\{ \int_{-f_2}^{f_1}\left( \sum_{j\geq
0}\frac{(-1)^j}{k_j} \frac{\p u_j}{\p x} (x,y)\right)
\sin\left(\frac{k_p(2y+f_2-f_1)}{f_1+f_2}\right)\d y \right\},
\end{subeqnarray}
and
\begin{subeqnarray}
A^{(2)}_{in}(x) & = & -\delta_{in} \left(\frac{\p P_i}{\p y}(x,f_1(x)) \right)^{-1}  \frac{\p Q_i}{\p y}(x,f_1(x)) ,\\
A^{(3)}_{im}(x) & = & 2(-1)^m k_i \ell_m \left(\frac{\p P_i}{\p y}(x,f_1(x)) \right)^{-1}\int_0^1 Y_m(x,z) \sin (k_n z)\, \d z ,\\
A^{(4)}_{ip}(x) & = & 2 (-1)^p k_i k_p \left(\frac{\p P_i}{\p
y}(x,f_1(x)) \right)^{-1} \int_0^1 Z_p(x,z) \sin (k_n z)\, \d z
,\\
B^{(1)}_{ni}(x) & = & -\delta_{ni} \left(\frac{\p Q_n}{\p y}(x,-f_2(x)) \right)^{-1}  \frac{\p P_n}{\p y}(x,-f_2(x)) ,\\
B^{(3)}_{nm}(x) & = & 2(-1)^m k_n \ell_m \left(\frac{\p Q_n}{\p y}(x,-f_2(x)) \right)^{-1}\int_0^1 Y_m(x,z) \sin (k_n z) \,\d z ,\\
B^{(4)}_{np}(x) & = & 2 (-1)^{p+1} k_n k_p \left(\frac{\p Q_n}{\p
y}(x,-f_2(x)) \right)^{-1} \int_0^1 Z_p(x,z) \sin (k_n z) \,\d z
, \\
C^{(1)}_{mi}(x) & = &
\frac{2(-1)^i}{k_i\ell_mY_m(x,1)(f_1+f_2)} \int_{-f_2}^{f_1}
\frac{\p P_i}{\p y}(x,y)
\cos\left(\frac{\ell_m(2y+f_2-f_1)}{f_1+f_2}\right)
 \d y ,\\
C^{(2)}_{mn}(x) & = &
\frac{2(-1)^n}{k_n\ell_mY_m(x,1)(f_1+f_2)}\int_{-f_2}^{f_1}
\frac{\p Q_n}{\p y}(x,y)
\cos\left(\frac{\ell_m(2y+f_2-f_1)}{f_1+f_2}\right)
 \d y ,\\
C^{(4)}_{mp}(x) & = & \frac{-2k_p
Z_p(x,1)}{\ell_mY_m(x,1)(f_1+f_2)}\int_{-f_2}^{f_1}
\cos\left(\frac{\ell_m(2y+f_2-f_1)}{f_1+f_2}\right)
\sin\left(\frac{k_p(2y+f_2-f_1)}{f_1+f_2}\right)
 \d y ,\\
D^{(1)}_{pi}(x) & = & \frac{2(-1)^i}{k_ik_p
Z_p(x,1)(f_1+f_2)}\int_{-f_2}^{f_1} \frac{\p P_i}{\p y}(x,y)
\sin\left(\frac{k_p(2y+f_2-f_1)}{f_1+f_2}\right)
 \d y , \\
D^{(2)}_{pn}(x) & = & \frac{2(-1)^n}{k_nk_p
Z_p(x,1)(f_1+f_2)}\int_{-f_2}^{f_1} \frac{\p Q_n}{\p y}(x,y)
\sin\left(\frac{k_p(2y+f_2-f_1)}{f_1+f_2}\right)
 \d y ,\\
D^{(3)}_{pm}(x) & = & \frac{-2\ell_mY_m(x.1)}{k_p
Z_p(x,1)(f_1+f_2)}\int_{-f_2}^{f_1}
\cos\left(\frac{\ell_m(2y+f_2-f_1)}{f_1+f_2}\right)
\sin\left(\frac{k_p(2y+f_2-f_1)}{f_1+f_2}\right)
 \d y
\cdot
\end{subeqnarray}
Similar numerical technique as the ones illustrated in section
\ref{illustration} can be used to solve \eqref{dual2} and obtain
the leading-order velocities in the general case given by
\eqref{vgeneral} and \eqref{w0general}.

\printfigures

\end{document}